\documentclass[runningheads]{llncs}
\usepackage[T1]{fontenc}

\usepackage{iftex}
\usepackage{color}
\usepackage{xcolor}
\usepackage{comment}
\usepackage{ascmac}
\usepackage{amsmath,amssymb}
\usepackage{booktabs}
\usepackage{multirow}
\usepackage{graphicx}
\usepackage{wrapstuff}
\usepackage{marvosym}
\usepackage{pifont}
\usepackage{ebproof}
\usepackage{diagbox}
\usepackage{tcolorbox}
\usepackage{sty/thesis}
\usepackage{caption}
\usepackage{textcomp}
\usepackage{listings}

\usepackage[noabbrev]{cleveref}
\renewcommand{\figurename}{Fig.}
\crefname{figure}{Fig.}{Figs.}
\Crefname{figure}{Figure}{Figures}
\crefname{section}{Section}{Sections}
\crefname{table}{Table}{Tables}
\crefname{appendix}{Appendix}{Appendices}

\makeatletter
\AddToHook{cmd/appendix/before}{\def\cref@section@alias{appendix}}
\makeatother

\begin{document}

\title{Graph Rewriting Language as a Platform for Quantum Diagrammatic Calculi}
\titlerunning{Graph Rewriting Language for Quantum Diagrammatic Calculi}

\author{Kayo Tei\inst{1}\orcidID{0009-0005-1598-8538} \and
	Haruto Mishina\inst{1}\orcidID{0009-0008-5740-3165} \and\\
	Naoki Yamamoto\inst{1}\orcidID{0000-0003-1106-4491} \and
	Kazunori Ueda\inst{1}\orcidID{0000-0002-3424-1844}}
\authorrunning{K. Tei et al.}
\institute{Department of Computer Science and Engineering, Waseda University\\
	\email{\{tei, mishina, yamamoto, ueda\}@ueda.info.waseda.ac.jp}}
\maketitle
\begin{abstract}
	Systematic discovery of optimization paths in quantum circuit simplification remains a challenge.
	Today, ZX-calculus, a computing model for quantum circuit transformation,
	is attracting attention for its highly abstract graph-based approach.
	Whereas existing tools such as PyZX and Quantomatic
	offer domain-specific support for quantum circuit optimization, visualization and theorem-proving,
	we present a complementary approach using LMNtal, a general-purpose hierarchical graph rewriting language,
	to establish a diagrammatic transformation and verification platform with model checking.
	Our methodology shows three advantages:
	(1) a direct and concise encoding of the ZX-calculus, where quantifiers simplify complex rule specification;
	(2) a verification framework using state-space exploration and model checking to analyze rewrite strategies; and
	(3) an open platform for strategic experimentation combining programmable syntax with interactive visualization.
	Through case studies, we demonstrate how our framework helps understand optimization paths and design new algorithms and strategies.
	This suggests that the declarative language LMNtal and its toolchain could serve as a new platform
	to investigate quantum circuit transformation from a different perspective.
\end{abstract}

\section{Introduction}
\padlU{The drive towards practical quantum computing is heightening the
    need for}
\newsentence{effective methodologies for
    circuit representation and optimization}.
The ZX-calculus \cite{picturing_quantum_2017} has emerged
\padl{
    as a powerful graphical language for this purpose.
    \padlU{It allows us to handle quantum circuits as
        equational theories over diagrams,
        which provides both a higher-level of abstraction
        (compared to standard, matrix-based formulation) and
        expressiveness coming from the ability to handle diagrams
        with an arbitrary number of wires}
    \cite{Kissinger_2014,tensors_bang_graphs,graphical_reasoning}.
    %
}
\padl{
    State-of-the-art dedicated tools have addressed
    \padlU{this new formalism}
    from different angles.
    Performance-oriented libraries like PyZX \cite{kissinger2020Pyzx} introduce sophisticated, built-in heuristics
    to simplify large-scale circuits efficiently.
    Interactive proof assistants like Quantomatic \cite{quantomatic} provide a formal environment
    for verifying the correctness of specific proof steps.
}

\padlU{%
    On the other hand, the connection between the dedicated graphical
    calculus and and other computing and programming paradigms
    including general-purpose declarative languages remains largely
    unexplored.}
This paper addresses this gap by proposing a new method
using a general-purpose graph rewriting language, LMNtal
\cite{ueda_lmntal_2009}, and its extension QLMNtal
\cite{mishina_introducing_2024}.
\padlU{This is inspired by the affinity of the
    data structures the ZX-calculus and (Q)LMNtal handle.
    The goal of the present work is not a competitor of
    dedicated tools for circuit optimization
    but a laboratory for exploring and validating strategies for
    quantum computing.}%

\padlU{LMNtal was born as an attempt to unify
    constraint-based concurrency (a.k.a. concurrent constraint programming)
    \cite{FGCS-SCP} and Constraint Handling Rules \cite{CHR}.
    It later turned out to serve as
    a ``unifying'' formalism of
    diverse computational formalisms including process calculi,
    the $\lambda$-calculus and
    Proof Nets \cite{ueda_encoding_2008,takyu_encoding_2025}.}
%
%
%
\newsentence{%
    The publicly available
    implementation of LMNtal \cite{lmntal-github}
    supports state-space search
    and model checking of graph rewriting that scales up to $10^9$
    states \cite{gocho_evolution_2011}
    and comes with a state space visualizer for
    non-large
    problems \cite{lmntal_2010}.
}
\padl{
    Along with the toolchain,
    our approach utilizes QLMNtal's support for
    \textit{quantified} pattern matching.
    This feature allows us to directly and declaratively express ZX-rules
    over an arbitrary number of components
    ($n$ links, $m$ nodes) within a single, formal rewrite rule.

    \padlU{The above-mentioned affinity between the two paradigms
        overcomes the procedural complexity inherent
        in other approaches based on general-purpose languages and tools.}
    It makes it easier and practical to model, execute, and formally
    analyze the behavior of user-defined rewrite strategies.
    By leveraging LMNtal's built-in support for state-space search and LTL
    model checking,
    we can formally answer questions about the properties of a rule set,
    such as termination and confluence,
    before \padlU{building it into} high-performance tools.
}
%



This research makes three foundational contributions:


\noindent
1. \textit{Bridging Declarative Programming and the ZX-Calculus} ---
We show direct
encoding of the ZX-diagrams and rules in LMNtal
and demonstrate that QLMNtal's quantifiers provide a direct and concise
encoding for ZX-rules
involving an arbitrary number of nodes and links.

\padl{%
    \noindent
    2. \textit{Rewrite Strategy Verification Framework} ---
    We show how state-space exploration and model checking
    can be applied to the design and analysis of
    rewrite strategies.
    Our case studies show how this approach can be used to verify
    optimization paths,
    validate the execution trace of intricate proof procedures,
    and explore the behavior of non-confluent rule sets to gain practical insights.
}

\noindent
3. \textit{Open Platform for Strategic Experimentation} ---
\padl{%
    We provide an extensible infrastructure
    that combines a programmable, quantified syntax with interactive
    state-space visualization,
    establishing a
    platform for the
    analysis of quantum optimization strategies.
}


The paper proceeds as follows:
Section~2 introduces LMNtal, a graph rewriting language, and
QLMNtal, a recent extension of LMNtal with quantified rules.
Section~3 is a brief introduction of the ZX-calculus.
%
Section~4 details our basic LMNtal implementation of the
ZX-calculus along with quantified rule examples.
Section~5 presents the state space exploration framework with
several case studies.
%
Sections~6 and 7 present related and future work and conclusion.


\section{LMNtal: A Hierarchical Graph Rewriting Language}
We
briefly explain the hierarchical graph rewriting
language LMNtal
\cite{ueda_lmntal_2009}.
\newsentence{Unlike many other formalisms of graph rewriting,}
LMNtal consists of (i) term-based syntax,
(ii) structural congruence on terms that provides interpretation of
terms as graphs, and (iii) small-step reduction relation,
\newsentence{in the style of standard programming language definitions}.
For lack of space, we \padlU{omit} detailed syntactic conditions of
(i) and introduce (ii) and (iii) informally,
\newsentence{\padlU{adapting} from the description of
    \cite{takyu_encoding_2025},
    \padlU{and leaving
        details to \Cref{sec:more_LMNtal}}.
    We also leave related work on hierarchical graph rewriting
    (\cite{drewes_hierarchical_2002} and many others)
    to \cite{takyu_encoding_2025}.}
A tutorial introduction to LMNtal can be found in \cite{LMNtal_tutorial}
and the full formal definition in \cite{ueda_lmntal_2009}.

\subsection{Overview of LMNtal}\label{section: lmntal_overview}
The syntax of LMNtal is given in \cref{fig:lmntal-syntax}, where
three
syntactic categories, link names (denoted by $X$), atom names
(denoted by $p$), and possibly empty membrane names (denoted by $m$),
are presupposed.

\begin{figure}[tb]
    \hrule
    \vspace{3pt}
    \small
    \[
        \begin{array}{l@{\quad}r@{~}c@{~}l}
            \textit{Process}          & P   & ::=                                                                                                                & \zero \bigm| p\paren{X_1 \pc \ldots \pc X_n} \bigm| P \pc P \bigm| m\mem{P} \bigm| T \react T \\[4pt]
            \textit{Process template} &
            T                         & ::= & \zero \bigm| p\paren{X_1 \pc \ldots \pc X_n} \bigm| T \pc T \bigm| m\mem{T} \bigm| T \react T \bigm| \texttt{\$} p
        \end{array}
    \]
    \hrule
    \vspace{3pt}
    \caption{Syntax of LMNtal.}
    \label{fig:lmntal-syntax}
    \vspace{-18pt}
\end{figure}

Since LMNtal was originally developed as a model of concurrency,
the hierarchical graphs of LMNtal are also called \textit{processes}.
$\zero$ is an inert process;
$p\paren{X_1 \pc \ldots \pc X_n} (n \geq 0)$ is an $n$-ary
\textit{atom} (a.k.a. node)
with \textit{ordered links} (a.k.a. edges) $X_1, \ldots, X_n$;
$P\pc P$ is parallel composition;
$\mem{P}$
is a \textit{cell} formed by wrapping $P$ with an
optionally named
\textit{membrane}
\verb+{+ \verb+}+; and
$T \react T$ is a \textit{rewrite rule}.

Occurrences of a link name represent endpoints of a one-to-one link
between atoms (or more precisely, atom arguments).
For this purpose, each link name in a process $P$ is
allowed to occur at most twice (Link Condition). A link whose name
occurs only once in $P$ is called a \textit{free link} of $P$. Links
may cross
membranes and connect atoms located at different ``places'' of the
membrane hierarchy.
A graph in which each node has its own arity and
totally ordered links,
like an LMNtal graph, is often called a \textit{port graph}
\cite{ene_attributed_2018}.

Process templates on both sides of a rewrite rule allow
\textit{process contexts}
\cite{uedalmntal_2005,ueda_lmntal_2009}.
A process context, denoted
$\verb+$+p$,
works as a
\textit{wildcard} that
matches ``the rest of the
processes''
within the membrane in which it
occurs.
%
\newsentence{%
    Whereas LMNtal allows us to specify what free links must occur in
    $\procvar p$, here we go without this feature, and $\procvar p$
    matches a process with any number of free links.}


Rewrite rules must observe several syntactic conditions
\cite{ueda_lmntal_2009} so that
Link Condition is preserved in the course of program execution.
Most importantly, link names in a
rule must occur exactly twice, and each process context must occur
exactly once at the top level of distinct cells in the left-hand side of a rule.

\begin{figure}[t]
    \centering
    \includegraphics[width=.95\columnwidth]{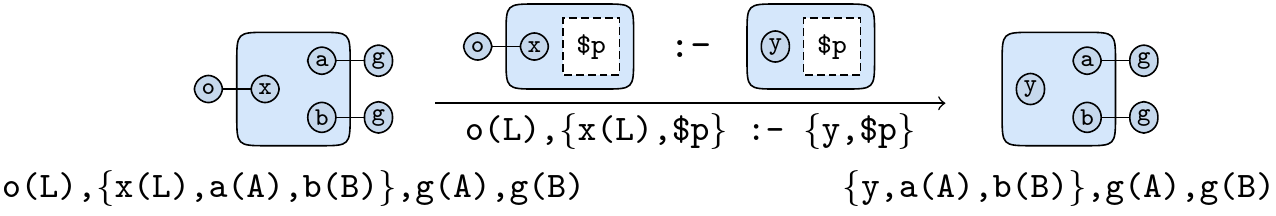}
    \caption{Rewriting with 
            {\small\texttt{(o(L),\{x(L),\$p\}) :- \{y,\$p\}}} .}
    \vspace{-1em}
    \label{fig:lmn_ex}
\end{figure}

\newsentence{%
    \Cref{fig:lmn_ex} shows a simple example of an LMNtal process
    containing a membrane, in both textual and visual forms,
    and a rewriting step using a rule containing a process context.}

The application of rules in LMNtal is nondeterministic because (i) a
rule may be able to rewrite different subgraphs of a given graph, and
(ii) different rules may be able to rewrite the same graph.
The LMNtal runtime SLIM
provides a \textit{nondeterministic execution
    mode} that constructs the whole state space
of rewriting, which
can also be visualized using
StateViewer \cite{lmntal_2010}.
Furthermore, SLIM provides an LTL model checker of the state space
\cite{gocho_evolution_2011}, and
\qcenew{%
    LaViT (LMNtal Visual Tools),
    which is a graphical IDE for LMNtal, provides
    visualizers of LMNtal graphs
    as well as a visualizer of state space called
    StateViewer.
}


LMNtal supports
\padlU{\textit{guarded} rewrite rules of the form
    ``$T \react \textit{Guard} \guard T.$'' }
Guards restrict rule applications based on data types, comparison
operators, and arithmetic operations.
In addition, we allow some abbreviations including:
\begin{itemize}
    \setlength{\itemsep}{0pt}
    \item \padlU{unary atoms `\texttt{+}' and
              `\texttt{-}' can be written without parenthesis as $\texttt{+X}$ and
              $\texttt{-X}$;}
    \item atoms in arguments are regarded as connected via `fresh' local links,
          e.g., $\texttt{a(b,c)} \equiv \texttt{a(B,C),b(B),c(C)}$;
    \item a reserved binary atom `\texttt{=}', called a
          \emph{connector}, fuses two links;
    \item
          a rule can be prefixed with a rule name followed by
          ``\texttt{@@}''.
\end{itemize}

\subsection{QLMNtal: LMNtal with Quantification}\label{subsec:qlmntal}

QLMNtal\footnote{%
    \padlU{The LMNtal runtime SLIM already fully supports intermediate code
        for QLMNtal, though some of the quantified rules are currently
        hand-compiled.}}
%
\cite{mishina_introducing_2024}
is an extension of LMNtal that introduces \emph{quantification}
\newsentence{to both}
pattern matching and rewriting.
The main features of QLMNtal are as follows:
(i) it introduces three kinds of quantifiers,
\textit{cardinality}, \textit{non-existence}, and
\textit{universal quantification} in a unified setting;
(ii) it allows mixed and nested use of multiple quantifiers, enabling
complicated quantification in a single rewrite rule; and
(iii) it introduces \textit{labelled} quantifiers to control
the (in)dependency between different quantifiers 
occurring
within a
rewrite rule.

\begin{figure}[tb]
    \centering
    \begin{minipage}{0.45\textwidth}
        \begin{screen}[4]
            \begin{minipage}{0.55\textwidth}\footnotesize
                \begin{verbatim}
{+X1,+Y1,+Z1},
h{+X1,+X2},a(X2),
h{+Y1,+Y2},b(Y2),
h{+Z1,+Z2},c(Z2)
				\end{verbatim}
            \end{minipage}\hfill
            \begin{minipage}{0.37\textwidth}
                \vspace*{-6pt}
                \includegraphics[height=44pt]{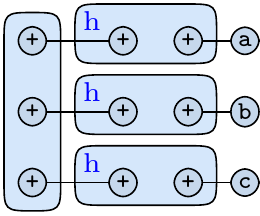}
                \hspace*{-20pt}
            \end{minipage}
            \vspace*{-9pt}
        \end{screen}
    \end{minipage}
    $\ \xrightarrow[\text{\small\texttt{zxh}}]{}\ $
    \begin{minipage}{0.32\textwidth}
        \begin{screen}[4]
            \begin{minipage}{0.65\textwidth}\footnotesize
                \begin{verbatim}
{+X1,+Y1,+Z1},
a(X1),
b(Y2),
c(Z3)
				\end{verbatim}
            \end{minipage}\hfill
            \begin{minipage}{0.23\textwidth}
                \vspace*{-6pt}
                \includegraphics[height=42pt]{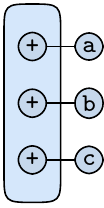}
                \hspace*{-20pt}
            \end{minipage}
            \vspace*{-9pt}
        \end{screen}
    \end{minipage}

    \begin{minipage}{0.4\textwidth}
        \begin{screen}[4]
            \begin{minipage}{0.55\textwidth}\footnotesize
                \begin{verbatim}
{+U,+V,+Z},
{+W,+X,+Y,+Z},
a(U),b(V),c(W),
d(X),e(Y)
				\end{verbatim}
            \end{minipage}\,
            \begin{minipage}{0.42\textwidth}
                \vspace*{-8pt}
                \includegraphics[width=\textwidth]{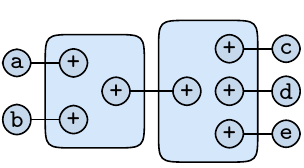}
            \end{minipage}
            \vspace*{-7pt}
        \end{screen}
    \end{minipage}
    $\ \xrightarrow[\text{\small\texttt{zxb}}]{}\ $
    \begin{minipage}{0.5\textwidth}
        \begin{screen}[4]
            \begin{minipage}{0.65\textwidth}\footnotesize
                \begin{verbatim}
{+U,+Z1,+Z2,+Z3},a(U),
{+V,+Z4,+Z5,+Z6},b(V),
{+W,+Z1,+Z4},c(W),
{+X,+Z2,+Z5},d(X),
{+Y,+Z3,+Z6},e(Y)
				\end{verbatim}
            \end{minipage}\,
            \begin{minipage}{0.32\textwidth}
                \vspace*{-6pt}
                \includegraphics[width=\textwidth]{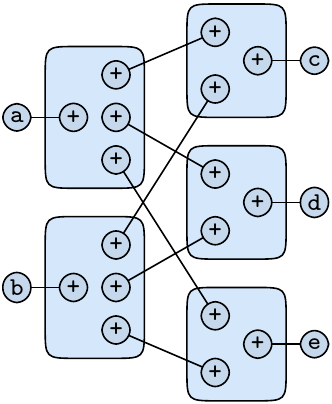}
            \end{minipage}
            \vspace*{-6pt}
        \end{screen}
    \end{minipage}
    \vspace*{-6pt}
    \caption{Rewriting with Rules \texttt{zxh} (upper) and
        \texttt{zxb} (lower).}
    \vspace*{-15pt}
    \label{fig:colour change in QLMNtal}
    \label{fig:bialgebra in QLMNtal}
\end{figure}

Cardinality quantification \newsentence{allows us to specify}
any number of processes within a
specified range and rewrite them in a single step.
For example,
\[\verb|zxh@@ {<*>+L1}, <*>h{+L1,+L2} :- {<*>+L2}|\]
is a QLMNtal rule that expresses the \padlU{simplified} Color Change rule
of the ZX-calculus
(\cref{subsec:Enabling Quantification in LMNtal}),
%
and \cref{fig:colour change in QLMNtal} shows an example of rewriting.
%
Here,
\verb|<*>| is a \textit{cardinality quantifier}
that represents ``zero or more'',
and the three cardinality quantifiers in this rule,
\newsentence{%
    related to each other by empty labels,
    stand for the same number of processes.}

QLMNtal also allows us to control the (in)dependency of quantifiers by
labeling. For example,
\[\small\verb|zxb@@ {M<*>+L1,+L2},{+L2,N<*>+L3} :- M<*>{+L1,N<*>+L2},N<*>{M<*>+L2,+L3}|\]
is a QLMNtal rule that expresses
\padlU{the essense of}
the (generalized) bialgebra rule (\cref{subsec:Enabling Quantification in LMNtal}).
\Cref{fig:bialgebra in QLMNtal} shows an example of rewriting.
\texttt{M} and \texttt{N} are \padlU{quantifier} labels; quantifiers
with different labels are independent of each other.
In this way, QLMNtal allows mixed and nested use of multiple quantifiers.
For
details including the syntax and formal semantics, the readers
are referred to \cite{mishina_introducing_2024}.

\section{ZX-Calculus}
Quantum programming frameworks such as
Qiskit \cite{ali_qiskit}
and
Cirq \cite{cirq} use circuit-based descriptions where
programs are expressed as temporally ordered sequences of operations
on qubits.
They
use basic
operations called \emph{quantum gates} to implement arbitrary quantum
computations. However, circuit representations alone often pose
challenges for optimization~\cite{nam2018automated,Amy2019} and
verification \cite{hietala2021sqir}.

The ZX-calculus \cite{CoeckeDuncan2011,picturing_quantum_2017} provides a
diagrammatic rewriting system using simple structures built from two types of
nodes connected by wires, serving as a computational model that
describes both quantum circuits and their transformations. This visual
and intuitive framework for representing quantum computations has a
rigorous mathematical foundation from \emph{string
	diagrams}~\cite{abramsky2009categorical,picturing_quantum_2017,biamonte2017tensor}.

\subsection{ZX-Diagrams}



The core component of the ZX-calculus, the ZX-diagram, consists of two
types of nodes: green (Z-spiders) and red (X-spiders), as in
\cref{fig:zx-spider}. Z-spiders correspond to the computational basis
$\{|0\rangle,|1\rangle\}$, whereas X-spiders correspond to the
Hadamard basis $\{|+\rangle,|-\rangle\}$. Each node possesses a phase
parameter $\alpha \in [0,2\pi)$ and any number of input/output
terminals.
\padl{Hadamard gates can be constructed from these spiders, also shown in
\cref{fig:zx-spider}.
}
The ``$\ldots$'' denotes \padlU{arbitrarily many}
wires.
\begin{wrapstuff}[type = figure, r, width = 0.35\textwidth]
	\centering
	\includegraphics[width=\columnwidth]{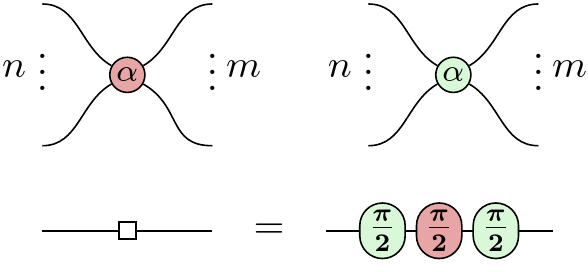}
	\caption{ZX-spiders (upper), Hadamard gate (lower) \cite{vandewetering_zx-calculus_2020}.}
	\label{fig:zx-spider}
\end{wrapstuff}

ZX-diagrams follow the principle of \emph{only connectivity matters}, meaning that wires can be arbitrarily bent or stretched as long as the connectivity is preserved.
Additionally, a single-input/\allowbreak output Hadamard gate can be formed using Z- and X-spiders as in \cref{fig:zx-spider}.
%
%
By connecting these components with wires, the framework provides a
unified representation of quantum gates, states, and protocols.
Further details of the foundations and the use of the ZX-calculus
can be found in
\cite{vandewetering_zx-calculus_2020,CoeckeDuncan2011,picturing_quantum_2017}.

\subsection{Basic Rewrite Rules}

\begin{figure*}[t]
	\centering
	\includegraphics[width=\textwidth]{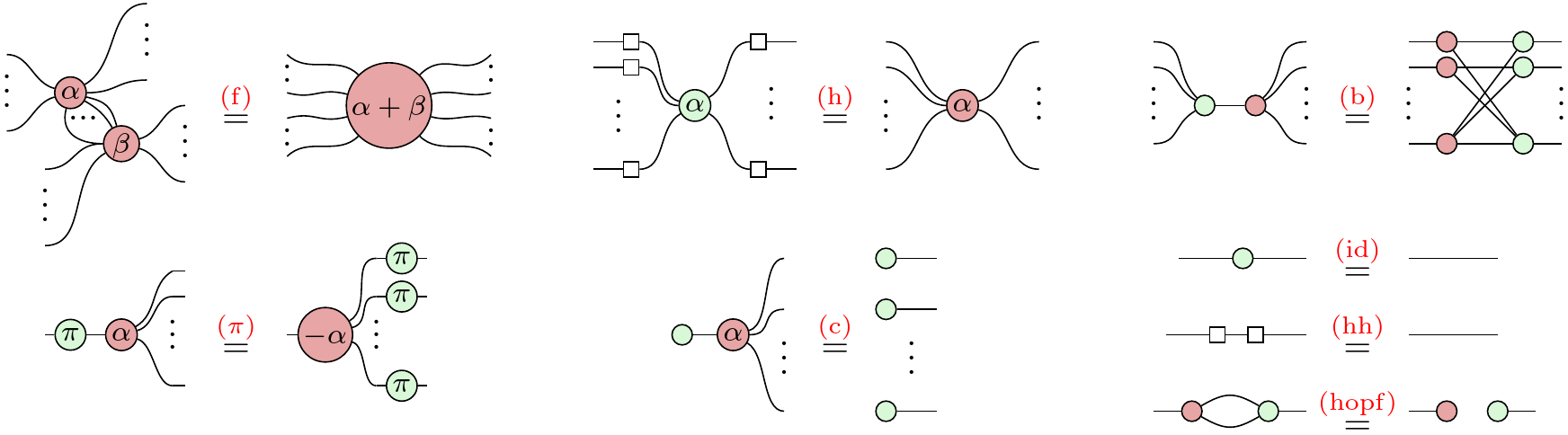}
	\caption{Key Rewrite Rules in ZX-Calculus \cite{vandewetering_zx-calculus_2020}. {\color{red}(f)}: spider fusion, {\color{red}(h)}: color change {\color{red}(b)}: (generalized) bialgebra rule, {\color{red}($\pi$)}: $\pi$ commutation, {\color{red}(c)}: copy, {\color{red}(id)}: identity, {\color{red}(hh)}: Hadamard cancellation, {\color{red}(hopf)}: Hopf.}\vspace*{-10pt}
	\label{fig:rules}
\end{figure*}

The ZX-calculus consists of rewrite rules of ZX-diagrams.
\padlY{\Cref{fig:rules} shows the key rules,}
\padlY{which allow} the \emph{color swapping} \cite{coecke2014zx}
symmetry (Z $\leftrightarrow$ X) and enable visual verification of
complex quantum protocols. For
details,
see
\cite{vandewetering_zx-calculus_2020,CoeckeDuncan2011,picturing_quantum_2017}.

\qcenew{
	\subsection{Circuit Extraction Problem}
	The conversion from quantum circuits to ZX-diagrams is straightforward by replacing each gate with its corresponding ZX-diagram. However, the reverse problem is known to be challenging. As mentioned in \cite{vandewetering_zx-calculus_2020}, there are two approaches to address this issue.

	The first approach is to extend the definition of quantum circuits to allow arbitrary linear maps between qubits,
	\padl{which requires extending the definition of a circuit to include a broader class of quantum operations.}
	In LMNtal, this can be achieved by defining a rule that translates each spider into a series of CNOT gates, which can then be applied to obtain the circuit.

	The second approach is to restrict ZX-diagrams to represent only
	unitary computations.
	Since unitarity is a global property of a diagram, it is challenging
	to construct circuits efficiently by examining their local
	structure only.
	%
	\padl{Recently, methods have been proposed that focus on a structural property of the diagram known as gflow} \cite{duncan-et-al-2020,there_back_again}.
	These methods utilize specific graph transformation rules,
	such as local complementation and pivoting (\cref{fig:lcomp-pivot}),
	to systematically reshape the diagram into a circuit-like structure.

	\begin{figure}[t]
		\centering
		\vspace*{-3pt}
		\includegraphics[width=0.65\columnwidth]{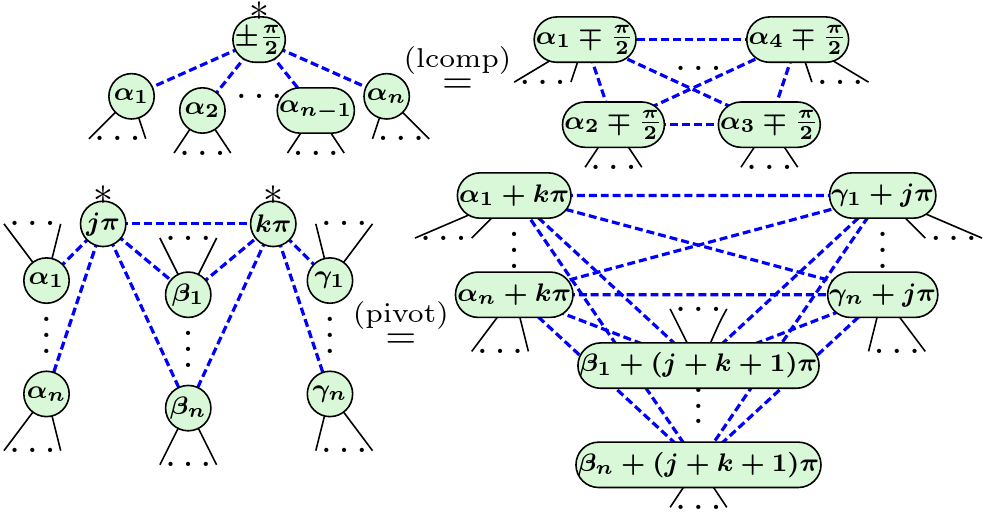}
		\caption{ZX-rules based on local complementation and pivoting. Blue dotted wires are wires with a Hadamard gate \cite{vandewetering_zx-calculus_2020}.}
		\label{fig:lcomp-pivot}
	\end{figure}
}

\qcenewU{
	\subsection{!-Boxes}\label{subsec:bang_boxes}
	The (informal) ``\dots'' notation is used
	to represent an arbitrary number of wires
	in ZX-diagrams, but a formal expression that avoids possible
	confusion has been proposed.
	The !-box \cite{
		Kissinger_2014,tensors_bang_graphs,graphical_reasoning} is represented as a square
	enclosing a part of the ZX-diagram, indicating that the enclosed part
	can be copied an arbitrary number ($\ge 0$)
	of times. Multiple !-boxes
	can exist within a single diagram, and they can overlap or be nested.
	%
	The LMNtal counterpart of the !-boxes is the quantifiers of QLMNtal,
	which
	allow nested, labelled quantifiers and allow one
	to express rules like (generalized) bialgebra, as was shown in
	\cref{subsec:qlmntal}.
	\padl{
	However, translating these powerful !-box diagrams into executable rewrite rules
	can be complex in many programming environments, often requiring procedural workarounds.
	}

	\padl{
		There appears to be a direct correspondence between the structure of !-boxes and QLMNtal's quantifiers. An !-box enclosing a part of a diagram can be systematically translated into a cardinality quantifier {\tt <*>} (zero or more) acting on the corresponding LMNtal atoms. Furthermore, in more complex scenarios, nested or overlapping !-boxes correspond to the nested, labeled quantifiers in QLMNtal.
The generalized bialgebra rule in \cref{fig:bang-bialgebra} serves as
a prime example. The two !-boxes map directly to the two quantifiers
({\tt M<*>} and {\tt N<*>}) in the QLMNtal rule \texttt{zxb}
\padlU{(\cref{subsec:qlmntal}),}
\padlU{where the nested quantifier on the right-hand slide
stands for multiplicative copies of the link \texttt{L2}}.
This
suggests the potential for a direct, systematic encoding from !-graph
rules to QLMNtal rules.

		\begin{figure}[t]
			\centering
			\includegraphics[width=0.9\columnwidth]{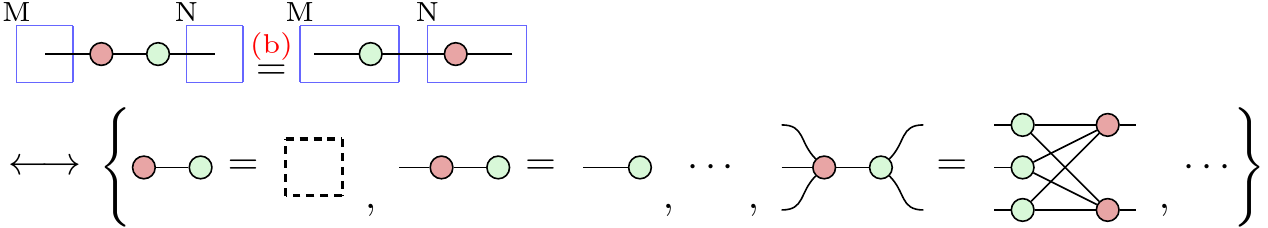}
\caption{!-graph for the generalized bialgebra rule
\padlU{{\color{red}(b)} (\cref{fig:rules})}.
  Each blue box is a !-box \cite[
    p.47]{vandewetering_zx-calculus_2020}.}\vspace*{-12pt}
			\label{fig:bang-bialgebra}
		\end{figure}
	}

	Although
	!-boxes are powerful, not all
	repetitive structures can be expressed.
	For example, they cannot represent
	complete graphs or
	local complementation
	\cite{vandewetering_zx-calculus_2020}, which is the case with QLMNtal also.
	%
	However, we have confirmed that QLMNtal allows one to express pivoting
	(a.k.a. edge-local complementation) as a single rewrite rule.  The
	idea is that, unlike complete graphs,
	it is sufficient to be able to express \textit{complete bipartite graphs}
	to express pivoting.
	Furthermore, as illustrated in \cref{sec:impl},
	LMNtal allows one to express rewriting strategies or procedures
	within the language, i.e., by using additional or modified rules.

}

\padl{\section{ZX-Calculus Implementation in QLMNtal}\label{sec:impl}}
This section describes the implementation strategy for the ZX-Calculus
in LMNtal \padl{and QLMNtal}.%
\footnote{\qcenew{
		All the implementations shown in \cref{sec:impl} and \cref{sec:examples} are available at {\tt  \url{https://github.com/lmntal/ZX-calculus}}.}}

\medskip
\padl{
	\subsection{ZX-diagram}
}
The two basic components of the ZX-diagram, spiders and wires, can be
directly represented using LMNtal membranes and links, respectively.
\qcenewU{Whereas each LMNtal atom
	has a fixed arity and its links are totally ordered,
	a membrane may hold
	outgoing links that are unordered.}
Thus, any ZX-diagram can be represented in LMNtal,
\qcenewY{since both frameworks
	do not care about diagram rotation or input/output.}
\padl{
	LMNtal's undirected links,
	which disregard length or orientation,
	ideally model the ``only connectivity matters'' principle of ZX-diagrams.
}
%

\begin{figure}[t]  
	\centering
	\includegraphics[width=0.6\columnwidth]{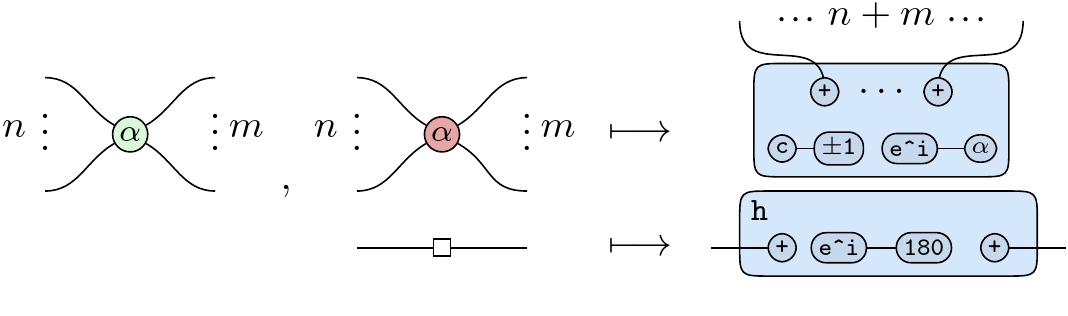}
	\caption{%
\padlU{Representation of Spiders (upper) and
and a Hadamard gate (lower).}}%
	\label{fig:spider_hadamard_lmn}
\end{figure}

\paragraph{Spiders.}
Although each membrane can be given a name,
we represent the two types of spiders with \textit{unnamed}
membranes as in \cref{fig:spider_hadamard_lmn}.
This is to treat the two types of spiders in a unified manner,
since
\padl{in}
the ZX-calculus,
\padl{inverting the color of each spider in a rule}
\padlU{gives us another valid rule.}
%
A membrane representing a spider holds
input/output wires, a \texttt{c} atom holding the
spider color,
and an \verb+e^i+ atom holding the phase
$\alpha\in[0,360)$.
%

\paragraph{Hadamard gates.}
\!\!Although Hadamard gates can be constructed from basic components,
we give them a dedicated LMNtal representation
(\cref{fig:spider_hadamard_lmn})
for simplicity.
A Hadamard gate is
represented using a membrane named \texttt{h},
which holds spider wires and a phase
\padl{equals to $\pi$}
in the membrane, but no atoms
representing colors.
For the reason why
\padlU{the \texttt{h} membranes hold the constant phase $\pi$,}
see \cref{sec:zh}.

\begin{figure}[tb]
	\centering
	\includegraphics[width=\columnwidth]{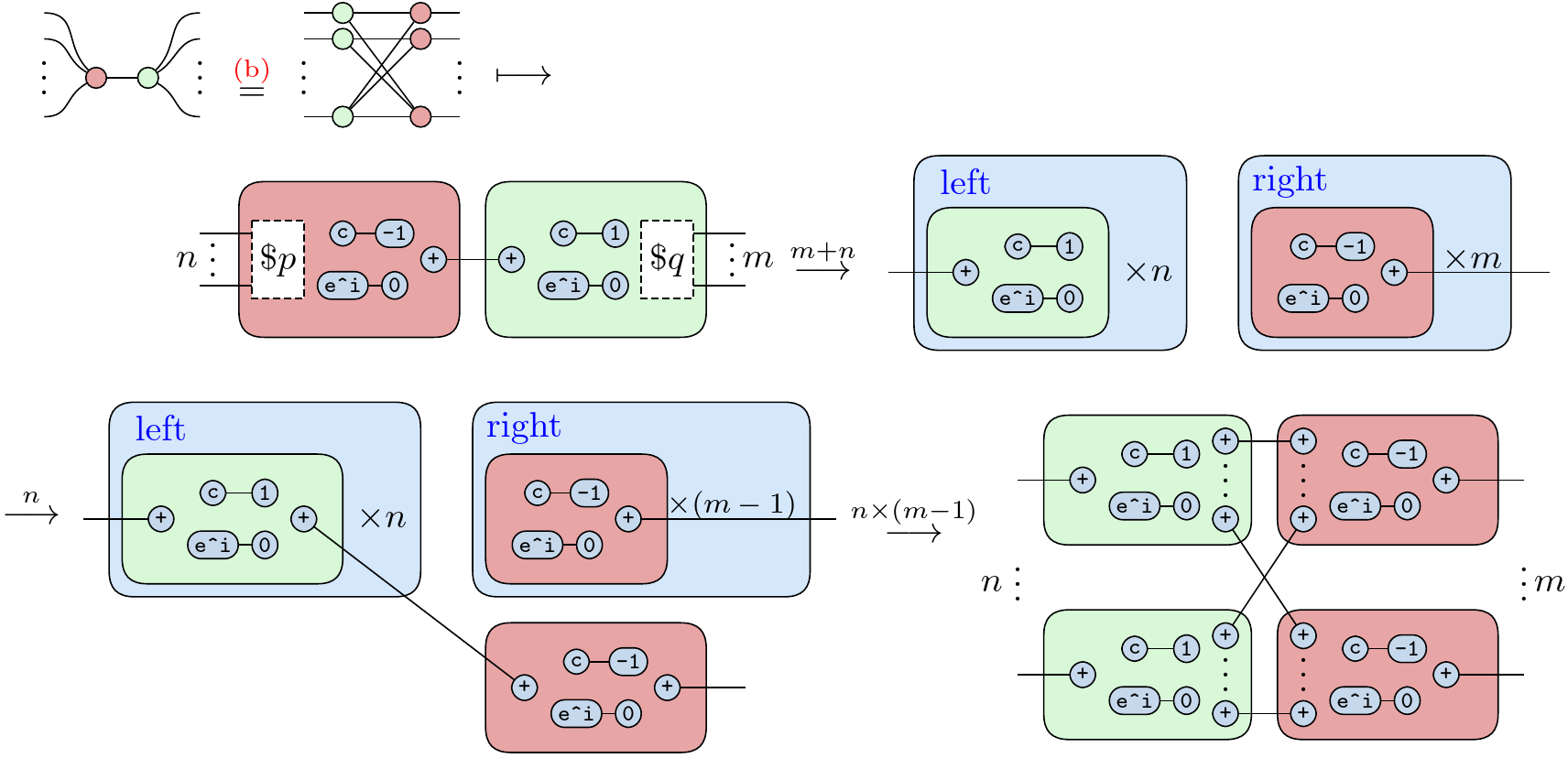}
	\caption{\padlU{Generalized bialgebra rule}
{\color{red}(b)} represented in plain LMNtal.}
	\vspace*{-10pt}
	\label{fig:bialgebra-complementarity}
\end{figure}

\padl{
	\subsection{The Challenge of Rule Implementation in plain LMNtal} \label{subsec:zx rule implementation}

	Plain LMNtal has sufficient expressiveness to implement
	the rewrite rules of the ZX-calculus.
	However, representing
	certain rules that handle an arbitrary number of wires,
	such as {\color{red}($\pi$)}, {\color{red}(c)}, {\color{red}(h)} and {\color{red}(b)},
	is not straightforward in plain LMNtal.
	These rules require a \emph{procedural} description consisting of multiple intermediate steps, rather than a single pattern-matching of diagrams.

	The complexity is most evident in the generalized bialgebra rule {\color{red}(b)}. This rule defines a operation that duplicates spiders according to the number of wires ($m$ and $n$, respectively) that are \textit{not} connected between two spiders of different colors, and connects them to form a complete bipartite graph.
	\padl{
Implementing this in plain LMNtal requires a multi-step procedure combining several rules:
	}%
\padlU{(i)}
counting the number of wires $m$ and $n$,
\padlU{(ii)}
duplicating the spiders
according to that number, and
\padlU{(iii)}
correctly connecting the duplicated spiders.
An overview of this \padlU{procedure}
is shown in \cref{fig:bialgebra-complementarity}.
	\padl{
Specifically, this implementation consists of 10 LMNtal rules and requires $O(mn)$ rewriting steps in total.
See \cref{sec:zx-rules-in-plain-lmntal} for details of how
\padlU{this approach}
works for all the ZX-rules
\padlU{including {\color{red}(b)}}.
	}

	\padl{
		While this procedural implementation translates the declarative nature of a single ZX-rule into an executable workflow,
		it introduces intermediate states that do not directly correspond to valid ZX-diagrams.
		However, this is a manageable trade-off.
		Our framework offers control mechanisms to treat a series of rules as an atomic transaction.
		Furthermore, the abstraction techniques in our toolchain allow these intermediate states to be omitted from the state space, enabling analysis to focus solely on valid ZX-diagrams.
	}

	Note that
	implementing rules for a fixed, constant number of wires is straightforward and can still be done in a single LMNtal rule.
	To streamline this process, we have developed a converter that automatically generates LMNtal code
	from a ZX-diagram and rule representation, available at\\
	\texttt{\url{https://github.com/lmntal/zx-lmn-converter}}.
}

\vspace*{-12pt}
\subsection{Simplifying the Encoding with QLMNtal}
\label{subsec:Enabling Quantification in LMNtal}


As shown in \cref{subsec:zx rule implementation}, the implementation
of ZX-rules
requires multiple LMNtal rules and thus multiple
execution steps to represent
	{\color{red}($\pi$)}, {\color{red}(c)}, {\color{red}(h)}, and
	{\color{red}(b)}.
However,
the cardinality construct introduced into QLMNtal
turns out to be powerful enough to express
each
ZX-rule
with a single QLMNtal rule as
in \cref{fig:ltr zx rule in QLMNtal}.
%
\qcenew{
	In particular, {\color{red}(b)} is implemented
	using nested quantifiers, as explained in
	\cref{subsec:qlmntal}.
}

\padl{
	The LMNtal implementations shown in \cref{fig:ltr zx rule in QLMNtal} are left-to-right transformations of the original ZX-rule equalities.
}
\qcenewY{
	This is because right-to-left applications of the rules may lead to an explosion of the state space in the non-deterministic execution of LMNtal.
	A right-to-left application of {\color{red}(c)} and {\color{red}(f)} 
	would generate
	arbitrary phases,
	and that of {\color{red} (id)} and {\color{red} (hh)} 
	would generate
	\padl{an identity spider}
	from an empty wire.}

\padl{While the right-to-left {\color{red}(b)} rule can be expressed in a single QLMNtal rule
and works,
	the inherent combinatorial complexity of the matching process at runtime remains.
}
\qcenew{
	However, if the user knows the number of wires needed for the rules,
	it is easy to add special
	cases of the right-to-left rules.
	Also, we can easily control
	how many times they should be applied
}\padl{%
	by adding any strategy.
	Our framework is thus positioned
\padlU{as a platform for formalizing and verifying}
	human-designed rewrite strategies.
	Users can define a set of directed rules and use our platform to exhaustively analyze its consequences, such as reachability, termination, and confluence for certain classes of diagrams.
}

\qcenew{%
	We would like to emphasize that in QLMNtal, all right-to-left
	transformation rules
	in \cref{fig:ltr zx rule in QLMNtal} can also be implemented with a
	single rule by \emph{simply swapping its left- and right-hand sides}
        and adjusting guard conditions.
}

\begin{table*}[t]
	\centering
	\caption{Left-to-Right QLMNtal implementation of ZX-rules.}
        \medskip
	\begin{tabular}{c|l}
		\hline
		ZX-rule              & \multicolumn{1}{c}{QLMNtal implementation} \\
		\hline
		\hline
		{\color{red}($\pi$)} &
		\begin{tabular}{l}
	\verb/{+L1,+L2,e^i(180),c(C1)}, {+L2,<*>+L3,e^i(A),c(C2)} :- / \\
	\verb/  AA=-A, C1*C2=:=-1 |/          \\
	\verb/  {+L1,<*>+L2,e^i(AA),c(C2)}, <*>{+L2,+L3,e^i(180),c(C1)}/ \\
		\end{tabular}
		\\
		\hline
		{\color{red}(c)}     &
		\begin{tabular}{@{}l@{}}
	\verb/{+L1,e^i(0),c(C1)}, {+L1,<*>+L2,e^i(A),c(C2)} :-/             \\
	\verb/  int(A), C1*C2=:=-1 | <*>{+L2,e^i(0),c(C1)}/ \\
		\end{tabular}
		\\
		\hline
		{\color{red}(h)}     &
		\begin{tabular}{@{}l@{}}
	\verb/{<*>+L1,e^i(A),c(C)}, <*>h{+L1,+L2,e^i(180)} :-/ \\
	\verb/  int(A), CC=-C | {<*>+L2,e^i(A),c(CC)}/  \\
		\end{tabular}
		\\
		\hline
		{\color{red}(b)}     &
		\begin{tabular}{@{}l@{}}
	\verb/{M<*>+L1,+L2,e^i(0),c(1)}, {+L2,N<*>+L3,e^i(0),c(-1)} :-/     \\
	\verb/  M<*>{+L1,N<*>+L2,e^i(0),c(-1)}, N<*>{M<*>+L2,+L3,e^i(0),c(1)}/ \\
		\end{tabular}
		\\
		\hline
	\end{tabular}
	\label{fig:ltr zx rule in QLMNtal}
\end{table*}

\section{Running the Examples}\label{sec:examples}

\begin{figure}[t]
		\centering
		\begin{minipage}[b]{0.42\columnwidth}
			\centering
			\includegraphics[width=\textwidth]{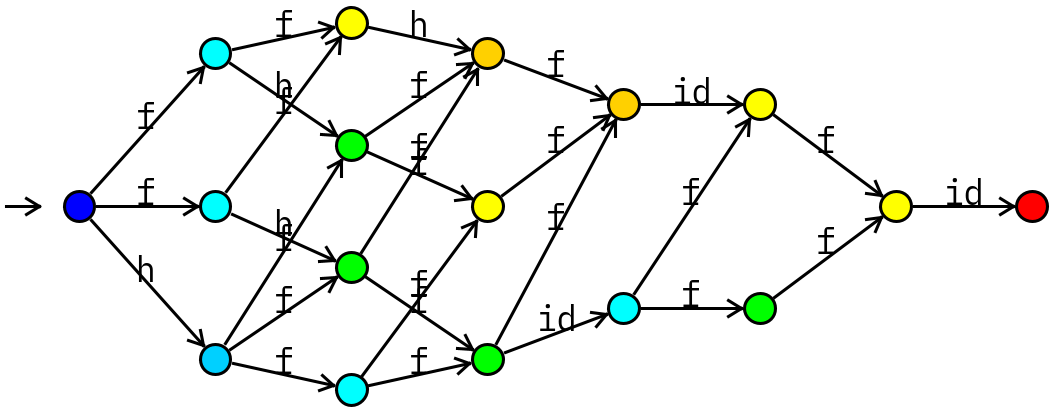}
			\caption{State space of the quantum teleportation circuit (17 states).}
			\label{fig:quantum_teleportation_state_space}
		\end{minipage}\hfill
		\begin{minipage}[b]{0.5\columnwidth}
			\centering
			\includegraphics[width=\textwidth]{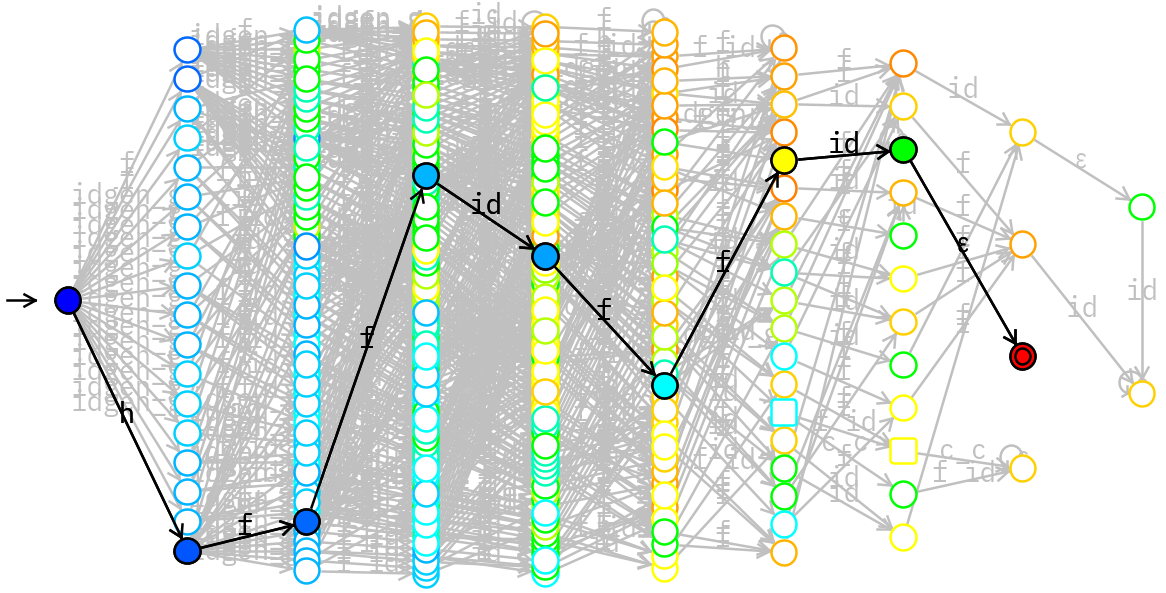}
			\caption{Application result of \texttt{idgen\_g} in the quantum
				teleportation circuit (313 states).}
			\label{fig:telep_idgen_g}
		\end{minipage}
	\end{figure}

\begin{figure}[t]
		\centering
		\begin{minipage}[b]{0.48\columnwidth}
			\centering
			\includegraphics[width=\textwidth]{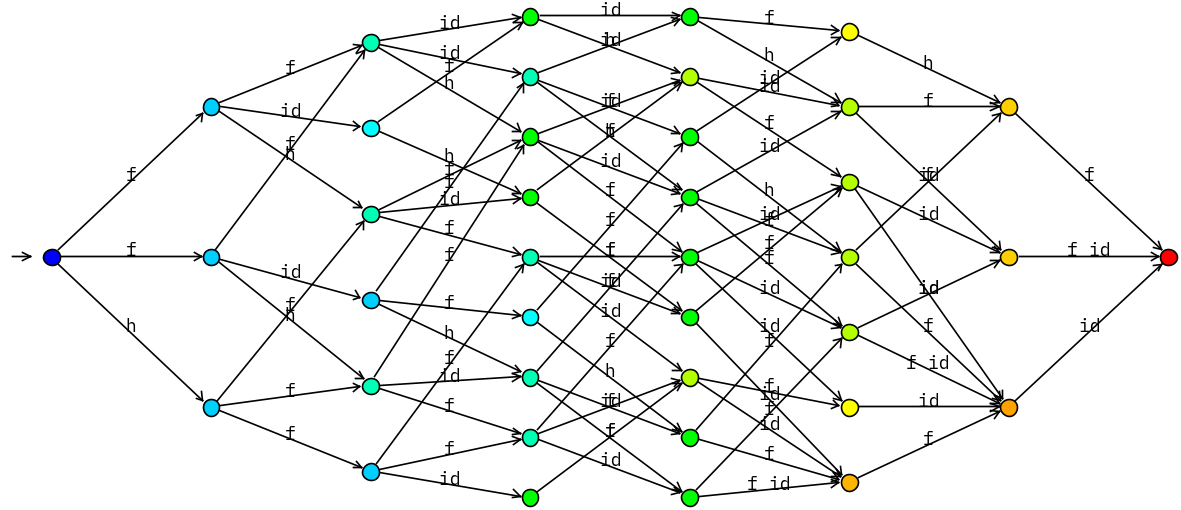}
			\caption{State space of the 3-bit GHZ preparation circuit (39 states).}
			\label{fig:3ghz_state_space}
		\end{minipage}\hfill
		\begin{minipage}[b]{0.48\columnwidth}
			\centering
			\includegraphics[width=\textwidth]{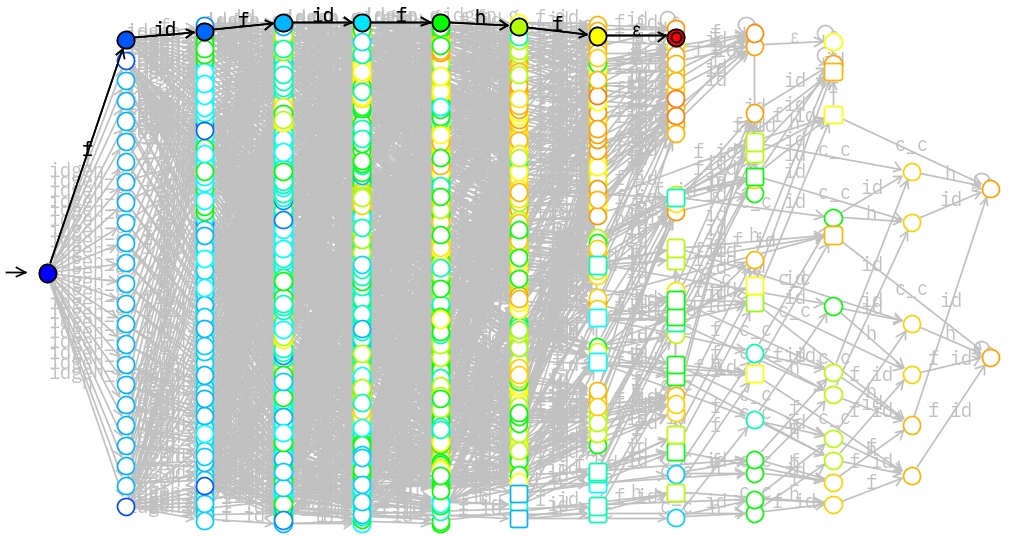}
			\caption{Application result of \texttt{idgen\_g} in the 3-bit GHZ preparation circuit (800 states).}
			\label{fig:3ghz_idgeng}
		\end{minipage}
		\vspace*{-9pt}
	\end{figure}

\padl{This section demonstrates}
our implementation using \qcenew{three} example problems.
All descriptions in this section are based on the ZX-calculus
implementation in
\padl{
	QLMNtal, showcasing how the platform can be used to model and verify different classes of rewrite strategies.
}
Details on the example problems
can be found in
\cite[Section 5]{vandewetering_zx-calculus_2020}.
\Cref{fig:quantum_teleportation_state_space,fig:telep_idgen_g,fig:3ghz_state_space,fig:3ghz_idgeng,fig:lcomp-state}
are state transition diagrams output using StateViewer
(\cref{section: lmntal_overview}),
\padlU{which shows the final states}
in red \qcenew{so they can be easily
	\qcenewU{identified} from the state space.
\padlU{LMNtal's runtime SLIM}
reports the number of final states and their corresponding graphs.
Additionally, \padlU{LaViT allow us to visualize}
the graph associated with each state.
}
%

\padl{
	\subsection{Analyzing Optimization Paths}\label{Analyzing Optimization Paths}

	\padlU{The} first case study demonstrates how our platform can
	be used
	to analyze optimization paths by introducing a simple,
	non-confluent rule and checking its impact on the state space.
	The goal is to answer a common question in
	optimization: \emph{is it ever beneficial to temporarily
	increase a diagram's complexity to find a better solution?}

	To model this, we introduce a right-to-left version of the {\color{red}(id)} rule,
	\texttt{idgen\_g}, which adds a Z-spider to an arbitrary wire.
	While potentially useful in some contexts, such a rule can also create inefficient pathways and dramatically expand the state space.
	We applied a basic set of simplification rules, {\color{red}(f)}, {\color{red}(id)}, and {\color{red}(h)},
	to two typical circuits,
        \padlU{(i)}
        the quantum teleportation
	circuit \cite{PhysRevLett.70.1895,CoeckeDuncan2011}
	and
        \padlU{(ii)}
        the GHZ preparation circuit, which
	generates the Greenberger-Horne-Zeilinger state \cite{Greenberger1989-GREGBB,kissinger-wetering-2020}.
	We then compared the resulting state space to
        \padlU{the} one where the \texttt{idgen\_g} rule was permitted
	to be applied at most once:
}

{\small
	\begin{verbatim}
  idgen_g@@ {+L1,$p1}, {+L2,$p2}, idgen_g(N)
    :- int(N), N>0, NN=N-1 |
    {+L1,+LL1,$p1}, {+L2,+LL2,$p2}, {+LL1,+LL2,e^i(0),c(1)}, idgen_g(NN).
\end{verbatim}
}

In order to perform
LTL model checking, we defined the proposition $p$
as
``\texttt{idgen\_g} is never applied and the
simplification has been completed,'' and by refuting the 
claim
that $p$ is never satisfied (\texttt{!<>}$p$), we showed that the path
where Z-spiders are not generated is the most efficient.

\begin{figure}[t]
		\centering
		\includegraphics[width=\columnwidth]{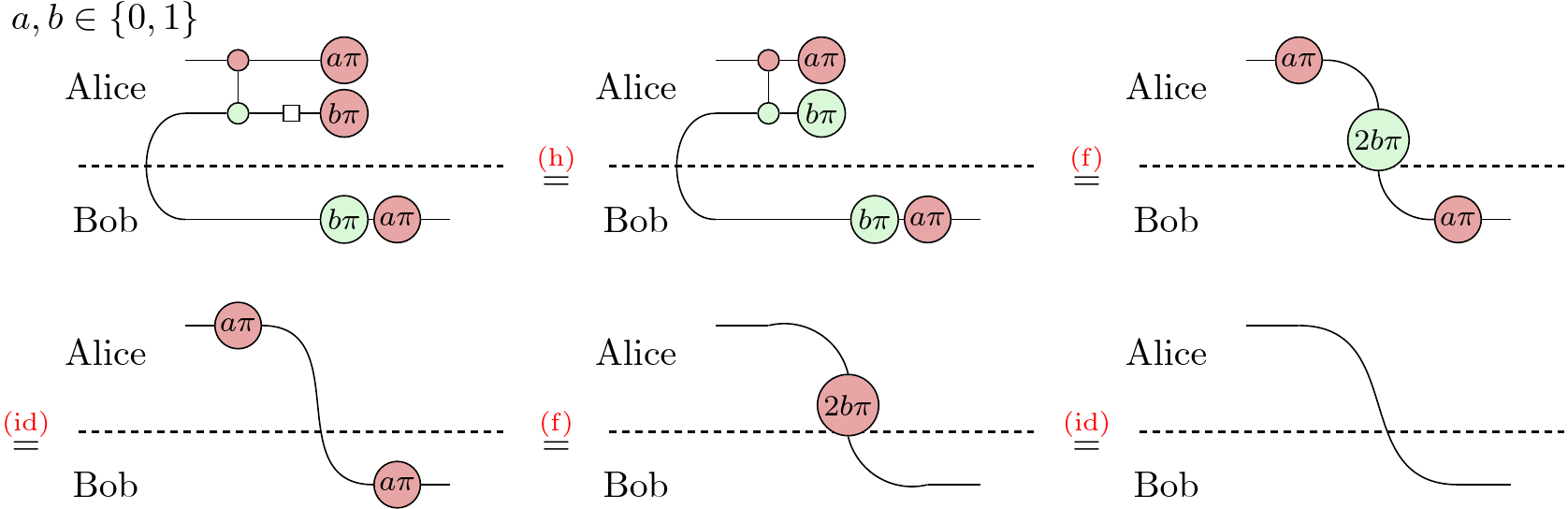}
		\caption{Simplification of the quantum teleportation circuit in ZX-calculus \cite{vandewetering_zx-calculus_2020}.}
		\vspace*{-10pt}
		\label{fig:quantum_teleportation}
	\end{figure}

	\paragraph{Quantum Teleportation Circuit:}
	An example of simplifying a quantum teleportation circuit is shown in \cref{fig:quantum_teleportation}.
	In this paper, we \newsentence{choose} fixed values for $a$ and $b$.
	The results for the quantum teleportation circuit are shown in \cref{fig:quantum_teleportation_state_space} (without \texttt{idgen\_g})
	and \cref{fig:telep_idgen_g} (with \texttt{idgen\_g}).

	We used LTL model checking to formally verify the property that ``the simplified final state can be reached without ever applying the \texttt{idgen\_g} rule.''
	The successful refutation of the claim \texttt{!<>}$p$ (where $p$ 
represents
completion without using \texttt{idgen\_g}) proves that all shortest paths avoid this rule.
	This confirms that, for this rule set, adding an identity spider only creates suboptimal solution paths.

\begin{figure}[t]
		\centering
		\includegraphics[width=\columnwidth]{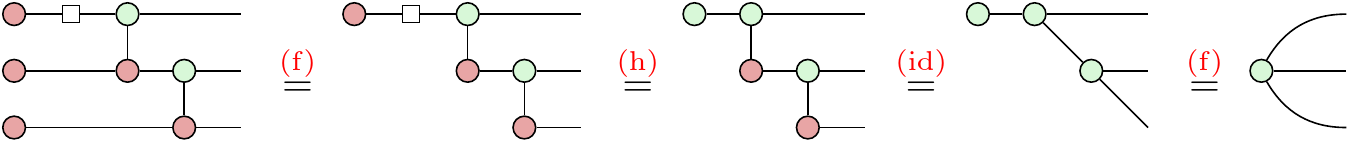}
		\caption{Simplification of the 3-bit GHZ preparation circuit \cite{vandewetering_zx-calculus_2020}.}
		\vspace*{-10pt}
		\label{fig:ghz_simplify}
	\end{figure}

\paragraph{GHZ Preparation Circuit:}
	The experiments on GHZ preparation circuits
	yielded the same conclusion.
	The 3-bit
	circuit can be simplified as in \cref{fig:ghz_simplify},
	and in \cref{fig:3ghz_state_space,fig:3ghz_idgeng}, we
	\padlU{show} examples of the experiment results on this circuit.
	This demonstrates the platform's ability to exhaustively analyze and validate the efficiency of a given optimization strategy.

\begin{wrapstuff}[type = table, r, width = 0.25\linewidth]
		\centering
		\small
		\caption{State space of GHZ preparation circuits.}
		\label{table:ghz3to10}
		\begin{tabular}{c|r}
			\hline
			Bit & \# of states \\
			\hline
			3   & 39           \\
			4   & 156          \\
			5   & 606          \\
			6   & 2424         \\
			7   & 9624         \\
			8   & 38596        \\
			9   & 153696       \\
			10  & 614784       \\
			\hline
		\end{tabular}
\vspace*{-12pt}
	\end{wrapstuff}

	We extended this analysis to GHZ preparation circuits from 4 to 10 bits.
	The results, shown in \cref{table:ghz3to10}, reveal an exponential growth pattern,
	where the number of states quadrupled for each additional qubit.
	For the 10-bit circuit, which starts from an initial graph of 28 spiders and one Hadamard gate,
	our system successfully explored all 614,784 states.
	Although the size of intermediate state graphs varies, considering LMNtal's demonstrated ability to explore state spaces of up to $10^{9}$ states,
	it is reasonable to assume that our framework can handle significantly larger problems.

\begin{figure}[t]
	\centering
	\includegraphics[width=\columnwidth]{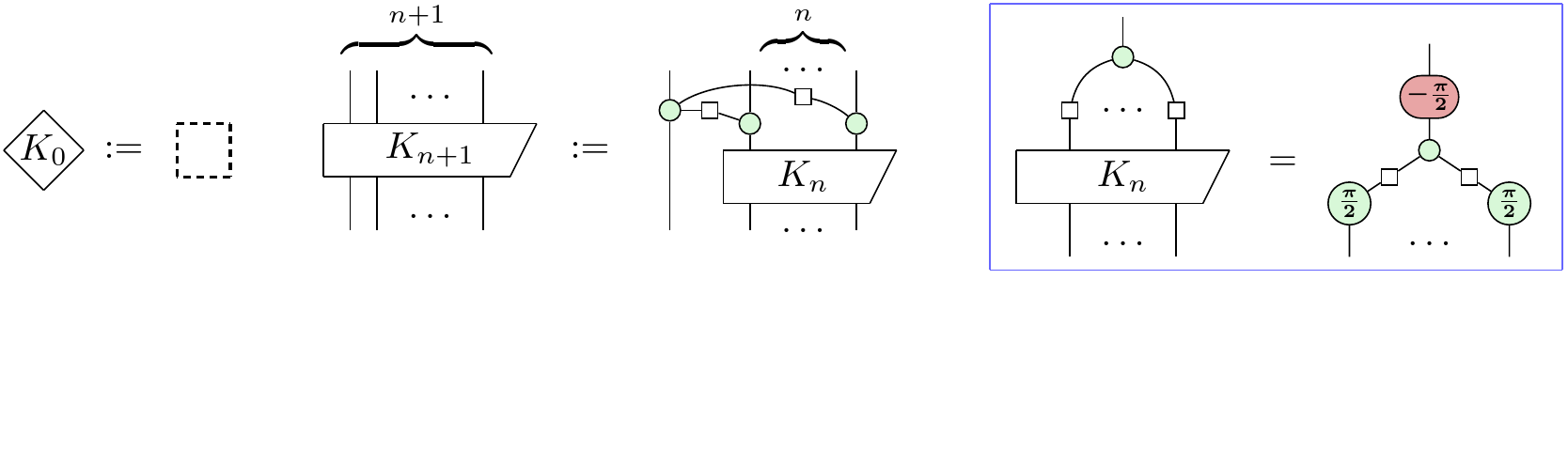}
	\vspace*{-56pt}
	\caption{Definition of $K_n$ and Lemma L (blue box) to be proved.}
	\vspace*{-10pt}
	\label{fig:lcomp-prob}
\end{figure}

\padl{
	\subsection{Verifying an Inductive Proof Strategy}\label{subsec:proof-lcomp}
	Beyond simple path analysis, our framework can formally verify complex, human-designed proof strategies for specific instances.
	We demonstrate this by encoding the inductive proof of a key lemma \cref{fig:lcomp-prob}
	from \cite[Lemma 9.129]{picturing_quantum_2017}, hereafter Lemma L, which is central to the local complementation rule.

	While the proof can be done inductively from ZX-rules as shown in \cite[Section 9.4]{picturing_quantum_2017},
	a model checker cannot handle
\padlU{mathematical}
induction over $n$, which is the domain of proof assistants.
	However, its strength
\padlU{is to be able to make sure}
that the logic of a proof holds for a concrete, non-trivial case,
\padlU{as}
an invaluable step
\padlU{towards}
validating a complex algorithm.
	Here, we
\padlU{check}
Lemma L for the $n=4$ case by checking the inductive step from $n=3$ to $n=4$.
}


\qcenew{%
	Lemma L is about a structure with a complete graph $K_n$ connected
	by Hadamard edges
	(\cref{fig:lcomp-prob}). In this
	procedure, first we \qcenewY{check} that Lemma L holds for
	$n=0$.
	\qcenewU{Then, we show that Lemma L also holds for $n=1,2,\dots,N$ for
		some $N$ using LMNtal rules representing the inductive definition.}
}

\begin{figure}[t]
	\centering
	\includegraphics[width=.95\columnwidth]{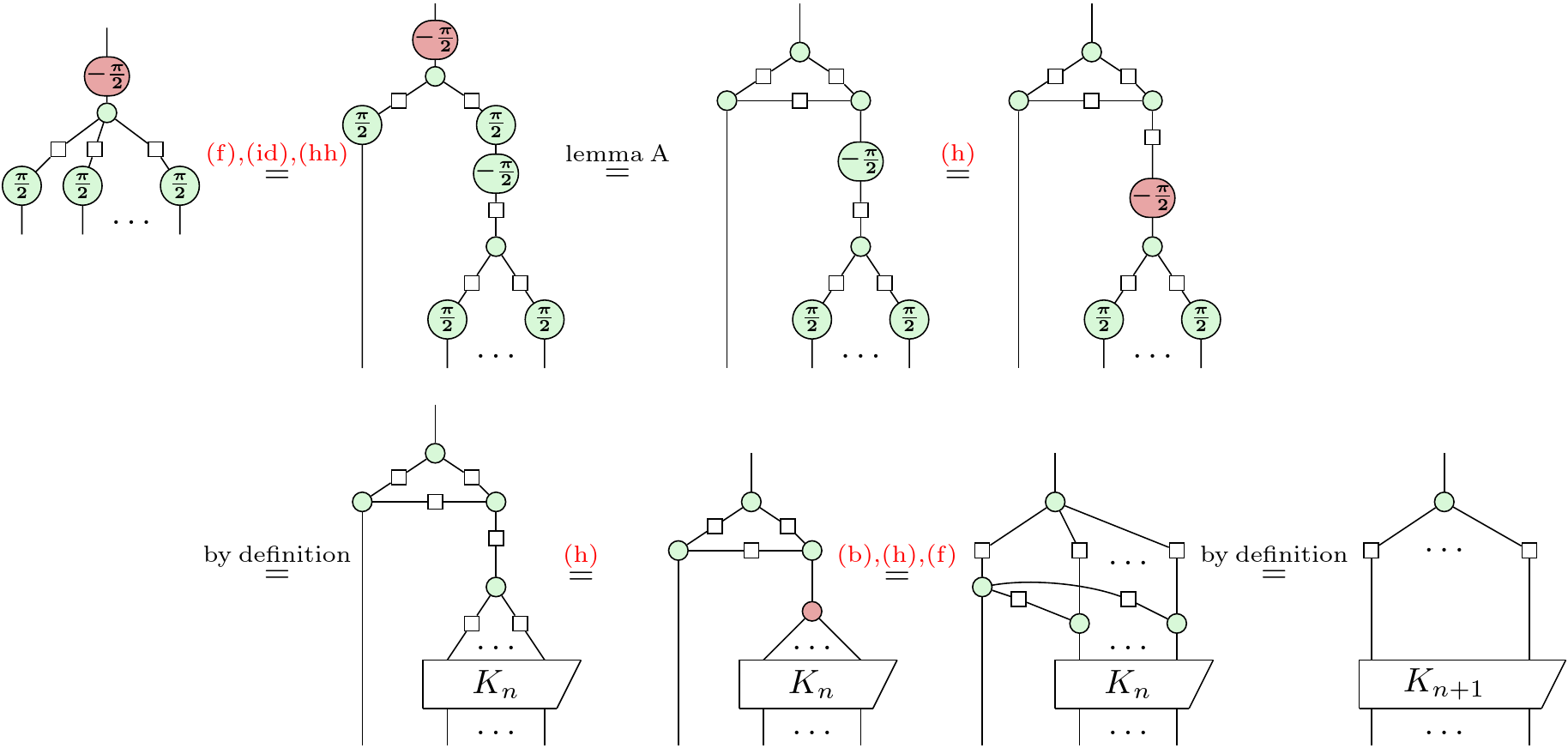}
	\caption{Proof Flow of Lemma L.}
	\vspace*{-10pt}
	\label{fig:lcomp-proof}
\end{figure}

\begin{wrapstuff}[type = figure, r, width = 0.5\textwidth]
	\centering
	\includegraphics[width=\columnwidth]{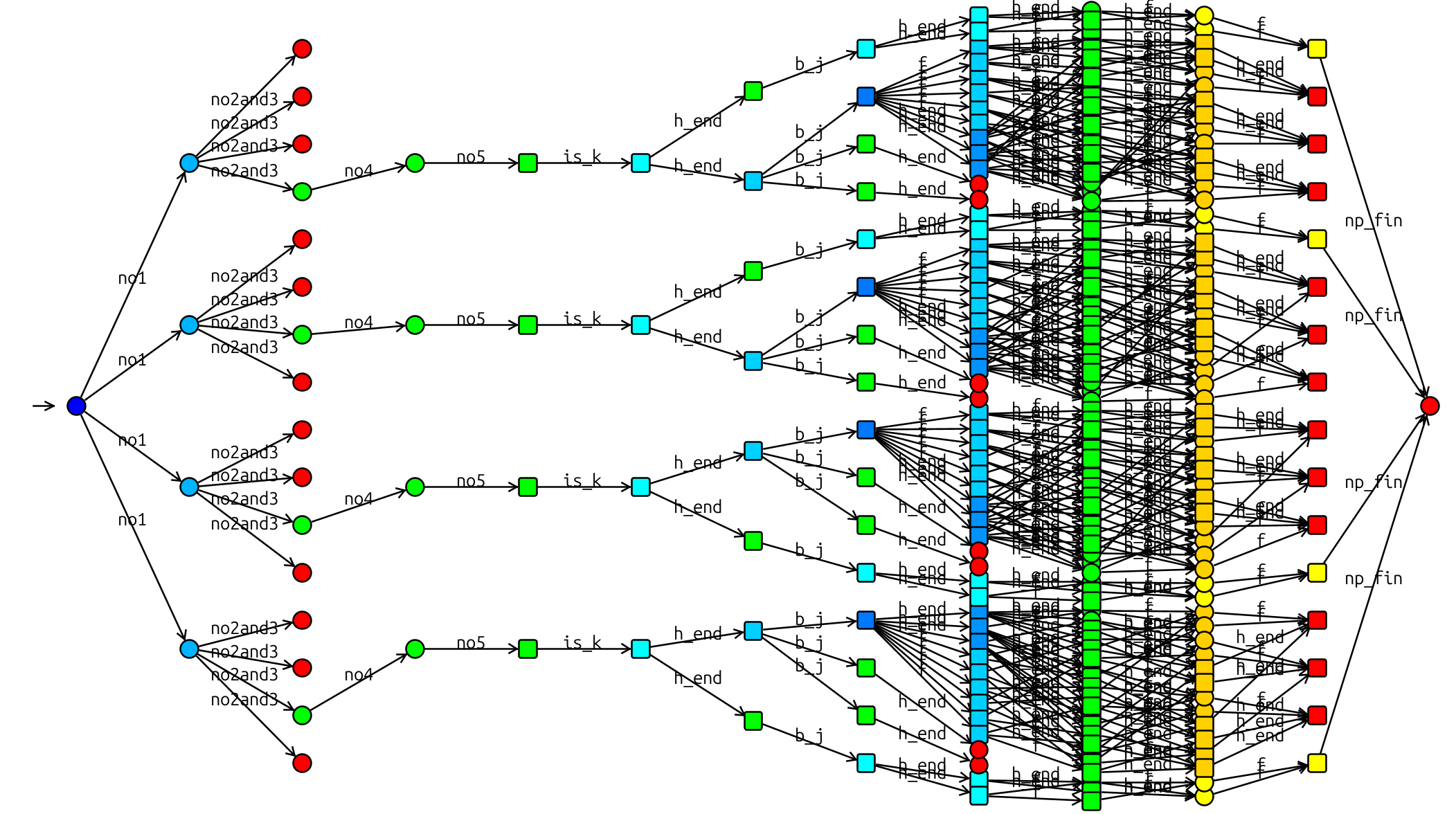}
	\caption{State Space of Lemma L ($n=4$, 266 states).}
	\label{fig:lcomp-state}
\end{wrapstuff}
\qcenew{
	The overall flow
	is shown in
	\cref{fig:lcomp-proof}. When the rewriting procedure is clear, we
	can avoid state space explosion by assigning a \textit{rule token} to
	each rule and preparing a \textit{linked list of the rule tokens}.
	\padl{
		This technique imposes procedural control onto the declarative rule set.
		A linked list of tokens, each corresponding to a specific rule in the proof sequence, is added to the graph.
		A rule is only permitted to fire if its corresponding token is at the head of the list.
		Upon firing, the rule consumes the token, activating the next rule in the sequence.
	}%
	In addition, for right-to-left rules, we can restrict the redexes
	by modifying the naively implemented ones.
	The state space generated by the $n=4$ case is
	shown in \cref{fig:lcomp-state}.
	\padl{
	The graph visualizes the proof trace itself.
	The initial branching shows the four possible choices for
	selecting the $n=3$ subgraph required by the inductive
	hypothesis.
	Each path then follows the encoded proof steps, all converging
	on the single final state on the far right, which is the
	expected conclusion of the lemma.
	This demonstrates how our platform can serve as a
\padlU{useful} tool for the mechanical verification of intricate,
	human-devised proof procedures.
	}
}


\padl{\subsection{Exploring Non-Confluent Strategies}\label{subsec:non-confluent}}
\padl{
	This case study demonstrates the platform's utility in exploring non-confluent state space.
	By analyzing the complete state space, we can gain insights into the rule set's behavior, identify ``dead ends'' (undesirable final states), and discover heuristics for guiding rewriting towards simplified forms.

	Our toolchain is well-suited for this analysis.
	The StateViewer visualizer allows for 
	interactive exploration of the paths leading to each terminal state,
	including tracing transitions backward and filtering them by rule names.
	This enables a detailed inspection of which rule sequences lead to desirable or undesirable outcomes.
	Furthermore, specific hypotheses about path efficiency can be formally specified and verified using LTL model checking.
}

\begin{wrapstuff}[type = table, r, width = 0.5\textwidth, vsep = 3pt]
	\centering
	\caption{\padl{State space of non-confluent examples}}
	\label{table:other_examples}
        \vspace*{-6pt}
	\small
	\begin{tabular}{c|r|r}
		\hline
		\begin{tabular}{@{}c@{}}
			Example \\
			(Rules)
		\end{tabular} &
		\begin{tabular}{@{}c@{}}
			\# of \\
			states
		\end{tabular} &
		\begin{tabular}{@{}c@{}c@{}}
			\# of \\
			final\\
			states
		\end{tabular}                \\
		\hline\hline
		\begin{tabular}{c}
			Pauli pushing \\
			({\color{red}($\pi$)}$\times 4$, {\color{red}(f)}, {\color{red}(id)})
		\end{tabular}
		                         & 533    & 7  \\\hline
		\begin{tabular}{c}
			2-qubit QFT \\
			({\color{red}($\pi$)}$\times 2$, {\color{red}(f)}, {\color{red}(id)}, {\color{red}(c)}, {\color{red}(h)})
		\end{tabular}
		                         & 1186   & 8  \\\hline
		\begin{tabular}{c}
			Detecting Entanglement \\
			({\color{red}(c)}, {\color{red}(f)}, {\color{red}(h)}, {\color{red}(hopf)}, {\color{red}(id)})
		\end{tabular}
		                         & 436711 & 60 \\
		\hline
	\end{tabular}
\end{wrapstuff}

The {\color{red}($\pi$)} and {\color{red}(b)} rules are recursive,
in the sense that 
their right-hand sides can match their left-hand sides, causing infinite loops.
To manage this,
we constrained their number of applications
to a
\padlU{constant}
sufficient for simplification in these experiments.
As shown in
\cref{table:other_examples}, we see that
\padl{%
the choice of the rule set impacts the scale of the strategic challenge.
The Pauli pushing example, with a relatively simple set of rules, generates a manageable but non-trivial state space with seven final states.
The Detecting Entanglement problem results in a larger state space, 
caused
by an interplay between specific rules.
The application of {\color{red}(hopf)}
can delete wires between spiders and create new graph structures that become valid targets for the {\color{red}(c)} rule.
Each application of {\color{red}(c)} then increases the number of spiders, thereby unlocking diverse rewrite paths.
}

From the state space by 2-qubit QFT,
we observed that applying the {\color{red} (id)} rule early in the rewriting process often led to ``dead ends'', and applying the {\color{red} ($\pi$)} rule in the middle stage tended to lead to more successful simplifications.
This suggests a testable heuristic for optimization:
rather than reducing the number of spiders in early stage,
applying certain rules to temporarily increase complexity may open up
more effective simplification paths later on.
Analyzing the conditions under which such heuristics hold in different
contexts is a promising direction for future
research using our platform.
For more details about the 2-qubit QFT example,
see
\cref{sec:qft}.

\bigskip
\section{Related Work} \label{sec:relatedwork}
\padl{%
    PyZX \cite{kissinger2020Pyzx}
    is a mature Python library
    optimized for large-scale quantum circuit simplification
    using built-in and learned heuristics.
    Our framework is not intended \padlU{to be} a
    competitor for
    optimization performance.
    Instead, it serves as a
    modeling platform
    to design and verify the underlying rewrite strategies themselves.
    For instance, a new simplification heuristic could be modeled
    in QLMNtal and its properties
    (e.g., confluence or termination for certain inputs) \padlU{be}
    analyzed in our system before being implemented in a
    performance-oriented tool like PyZX.%
}%

\qcenewU{%
    \padl{%
        Quantomatic \cite{quantomatic}
        is a graphical proof assistant for string diagrams,
        focusing on interactive, step-by-step theorem proving using
        \emph{directed} equations.
        Our approach is complementary, centered on the automatic,
        exhaustive state-space exploration of rewrite strategies using model checking.
        Whereas Quantomatic has graphical syntax for string diagrams,
        LMNtal is a general-purpose language with
        \padlU{programming language-style} text syntax
        and semantics supporting hierarchical port graphs
        \cite{LMNtal_tutorial}, which offers
        familiarity and more control (\cref{sec:examples}) for developers
        wishing to script and manage rule sets.
    }%
    Furthermore,
    the notation of hierarchy and wildcards in
    QLMNtal
    is
    expressive enough to capture
    the structures of
    string diagrams \padlU{in symmetric monoidal categories}.
}%

\qcenewU{%
    Comparison between LMNtal and other graph transformation languages and
    tools, including Groove and GP 2, can be found in
    \cite{mishina_introducing_2024} and \cite{takyu_encoding_2025}.%
}%

\section{Conclusion and Future Work}\label{sec:conclusion}
\padl{%
	This paper introduced
        \padlU{an approach and a tool}
	for modeling and analyzing
        complex rewriting strategies for quantum graphical calculi.
	We demonstrated that QLMNtal's quantified pattern matching
	can directly and declaratively express ZX-rules involving arbitrary numbers of wires, which can be non-intuitive and verbose
        \padlU{in} conventional procedural encodings.
	Using this representation,
	we presented a method based on state-space exploration and
	model checking to formally verify properties of human-designed rewrite strategies.
}

\padl{
	By combining
	quantified hierarchical graph rewriting
	with
	interactive state visualization,
	our framework serves as a laboratory for exploring new ideas in quantum computing.
	As shown in our case studies, this allows us to analyze optimization paths,
	validate the execution trace of intricate proof procedures,
	and explore the behavior of non-confluent rule sets to gain practical insights.
	This approach bridges the gap between the informal notation
	of diagrammatic reasoning and its executable, verifiable application.
}

\padl{
	For future work,
	we plan to extend our
	platform
	to other calculi
	like the ZW-calculus \cite{amar_zw}
	to build a comprehensive library of verified strategies.
	(We already implemented another variation, ZH-calculus \cite{backens2019zh}; see \cref{sec:zh}.)
	We also plan to systematically analyze existing optimization techniques
	to uncover new findings and utilize LMNtal's 
        capabilities 
        for novel global rewriting methods.

	On the practical side,
	extending our current translator to support input from
	standard formats like OpenQASM \cite{cross2017openqasm}
	would further enhance the platform's utility for the quantum computing community.
	Our team is also currently working on introducing probabilities
	into the LMNtal language and system,
	which we hope can lead to
	the modeling of measurement-based quantum computing \cite{zx-mbqc}.}

\clearpage
\bibliographystyle{splncs04}
\bibliography{ref}

\clearpage
\appendix 

\section{More details about the LMNtal language}\label{sec:more_LMNtal}

\newcommand{\equals}{\mathop{\texttt{=}}}
\newcommand{\reduces}{\longrightarrow}
\newcommand{\reducesR}[1]{\xrightarrow{\,#1\,}}
\newcommand{\narrowdots}{.\kern1pt.\kern1pt.\kern1pt}

For the readers new to LMNtal, this appendix
gives some more
details of LMNtal not in the main text.
Note that some papers handle processes and rewrite rules separately
(e.g., \cite{mishina_introducing_2024}), but this paper
follows the original definition \cite{ueda_lmntal_2009} and handles rewrite
rules as part of a process so that they can be placed inside membranes
to express local rewrite inside them.
\qcenewU{%
	We also briefly mention how we extended the semantics of LMNtal to
	incorporate quantifiers of QLMNtal.}


\subsection{Notes on the Syntax of LMNtal}\label{sec:syntax_LMNtal}

\begin{enumerate}
	\item
	      \textit{Parallel composition} $P_1,P_2$
	      glues two processes $P_1$ and $P_2$ to build a larger process.
	      Note that, if each of
	      $P_1$ and $P_2$ has a free link with the same name, it becomes a local
	      link in $(P_1, P_2)$.
	      A reader may notice that the Link Condition may not
	      always allow us to form $(P_1, P_2)$ from
	      two graphs $P_1$ and $P_2$ each satisfying the Link Condition.
	      How LMNtal handles this point will be discussed in
	      \cref{sec:SC_LMNtal}.

	\item
	\begin{wrapstuff}[type=figure,r,width=0.4\textwidth,vsep=6pt]

		      \centering
		      \includegraphics[width=9em]{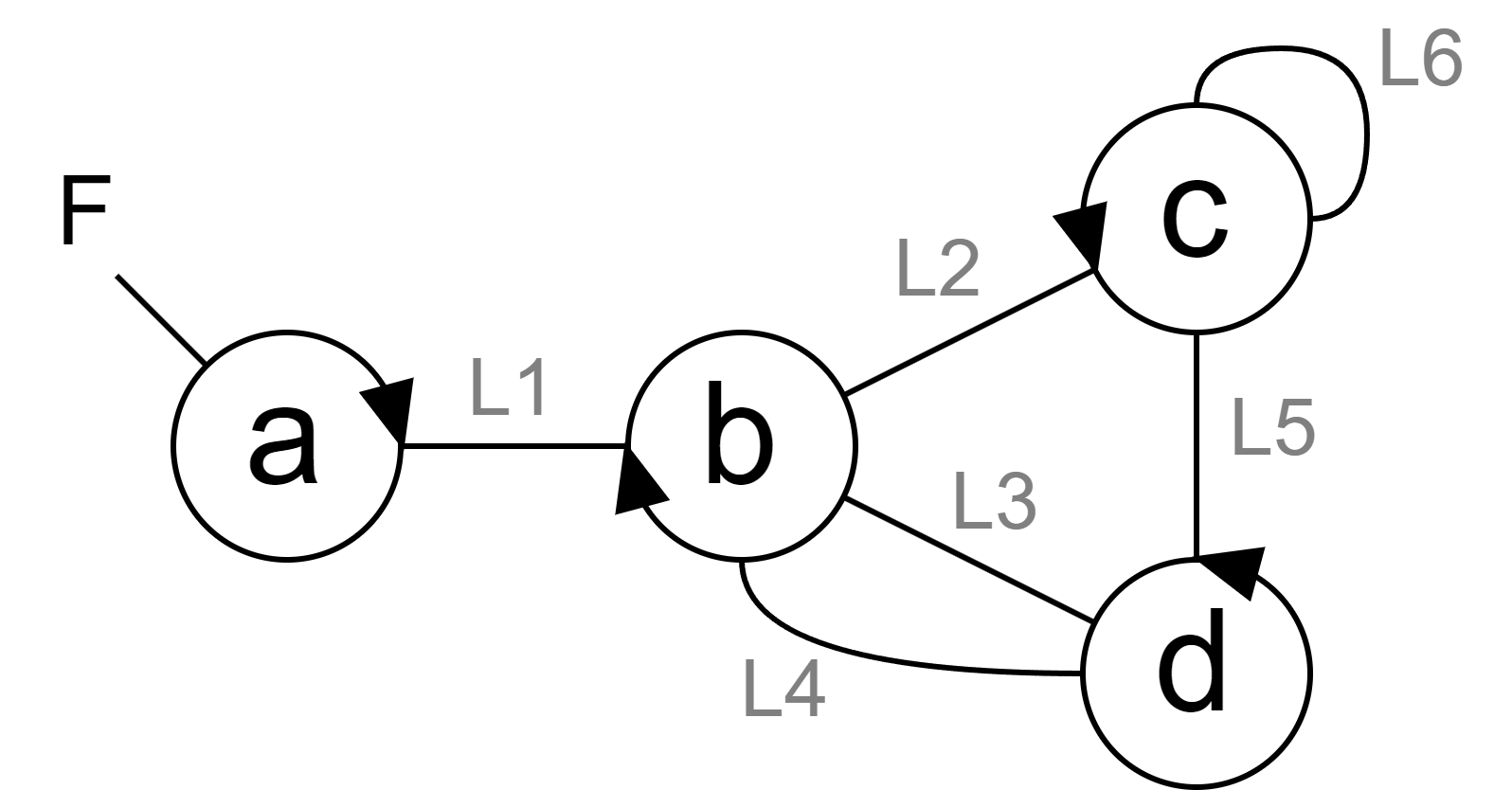}
      \caption{Pictorial representation of an LMNtal \padlU{(port)} graph,
		      in which \texttt{F} is a free link and the
                      others are local links.
                      \padlU{An arrowhead indicates the first
                      argument of an atom and the ordering of arguments.}}
		      \label{fig:lmntalexample}
	      \end{wrapstuff}
	      For readability, parallel composition may be written in a
	      period-terminated form as well as in a comma-separated form.  For instance,
	      \verb+a, (a:-b,c)+ may be written as \verb+a. a:-b,c.+\,,
	      where the comma binds tighter than `\texttt{:-}' which in turn
	      binds tighter than periods.

	\item
	      A special binary atom, called a \textit{connector} $\texttt{=}(X,Y)$,
	      also written as $X\equals Y$,
	      fuses (or glues) two links $X$ and $Y$.

	\item As a practical extension to the original definition
	      \cite{ueda_lmntal_2009}, the present syntax and our
	      implementation allow
	      named membranes of the form $m\mem{\,}$ (as in
	      \cref{section: lmntal_overview}) and
	      named rules of the form
	      $\textit{name}\,\texttt{@@}\ T\react T$.

	\item
	      A term representing a process is subject to Structural
	      Congruence
	      defined in \cref{sec:SC_LMNtal},
	      which then stands for an \textit{undirected port
	      multigraph}, i.e., a graph that allows
	      multi-edges and self-loops.
	      For instance, a process
	      \begin{center}
		      \texttt{a(L1,F),b(L1,L2,L3,L4),\\c(L2,L5,L6,L6),d(L5,L3,L4)}
	      \end{center}
	      stands for the undirected graph shown
	      in \cref{fig:lmntalexample},
	      where the order of atoms and the associativity of
	      `\texttt{,}' for parallel composition
	      are not significant because of the structural
	      congruence described in \cref{sec:SC_LMNtal}.
\end{enumerate}

\subsection{Semantics of LMNtal}\label{sec:semantics_LMNtal}

The semantics of LMNtal consists of \textit{structural congruence}
(\cref{sec:SC_LMNtal}) and a \textit{reduction relation}
(\cref{sec:reduction_relation}).

\subsection{Structural Congruence}\label{sec:SC_LMNtal}

The syntax defined in the main text does not yet characterize LMNtal graphs
because the port graph of \cref{fig:lmntalexample} allows
other syntactic representations.
\Cref{fig:Structural congruence on LMNtal processes}
defines an equivalence relation,
called
\textit{structural congruence}, to absorb the syntactic variations.
$P[Y/X]$ stands for the renaming of link $X$ in process $P$ to $Y$.
The rules apply only when each process satisfies the Link Condition.
The rules are as in \cite{ueda_lmntal_2009} and readers familiar with
LMNtal may skip the details.

\begin{figure}[tbp]
	\centering\small
	\begin{tabular}{rr@{~~}c@{~~}l}
		\hline                                                                              \\[-8pt]
		(E1)     & $0,P$                  & $\equiv$      & $P$                             \\[1pt]
		(E2)     & $P,Q$                  & $\equiv$      & $Q,P$                           \\[1pt]
		(E3)     & $P,(Q,R)$              & $\equiv$      & $(P,Q),R$                       \\[1pt]
		(E4)     & $P$                    & $\equiv$      & $P[Y/X]$                        \\
		\multicolumn{4}{r}{if $X$ is a local link of $P$}                                   \\[1pt]
		(E5)     & $P\equiv P'$           & $\Rightarrow$ & $P,Q\equiv P',Q$                \\[1pt]
		(E6)     & $P\equiv P'$           & $\Rightarrow$ & $m\mem{P}\equiv m\mem{P'}$      \\[1pt]
		(E7)     & $X\equals X$           & $\equiv$      & $\textbf{0}$                    \\[1pt]
		(E8)     & $X\equals Y$           & $\equiv$      & $Y\texttt{=}X$                  \\[1pt]
		(E9)     & $X\equals Y,P$         & $\equiv$      & $P[Y/X]$                        \\
		\multicolumn{4}{r}{if $P$ is an atom and $X$ is a free link of $P$}                 \\[1pt]
		(E10)    & $m\mem{X\equals Y,P}$  &
		$\equiv$ & $X\equals Y, m\mem{P}$                                                   \\
		\multicolumn{4}{r}{\hspace{5em}if exactly one of $X$ and $Y$ is a free link of $P$} \\[4pt]
		\hline
	\end{tabular}
	\caption{Structural congruence on LMNtal processes.}
	\vspace*{-10pt}
	\label{fig:Structural congruence on LMNtal processes}
\end{figure}

(E1)--(E3) characterize atoms as multisets.
(E4) is  $\alpha$-conversion of local link names, where $Y$ must be a
fresh link because of the Link Condition.
When we compose two graphs with a comma, we regard the two graphs to
be $\alpha$-converted by (E4) as necessary to avoid collisions of
local link names\footnote{When each of $G_1$ and $G_2$ satisfies the
	Link Condition but their composition
	$G_1 \pc G_2$ does not, there must be a link occurring twice in one
	and at least once in the other.  Since the former is
	a local link, it can be renamed by (E4) to restore the Link Condition.}.
(E5) and (E6) are
structural rules to make $\equiv$ a congruence.
(E7)--(E10) are rules for connectors:
(E7) says that a self-closed link is regarded as a empty graph;
(E8) states the symmetry of connectors;
(E9) says that a connector may be absorbed or emitted by an atom; and
(E10) says that connectors may cross membranes.


\qcenewU{%
	Structural congruence of QLMNtal is extended to
	express \textit{unrolling (one-step expansion)} of quantified
	processes \cite{mishina_introducing_2024}.}

\subsection{Term Notation}

For convenience, the following term notation is allowed and is widely
used; that is, we allow
%
\[
	p(X_1,\narrowdots,X_{k-1},L,X_{k+1},\narrowdots,X_m ),\,q(Y_1,\narrowdots,Y_{n-1},L)
\]
$(1 \leq k \leq m,1 \leq n)$ to be written as
\[p(X_1,\narrowdots,X_{k-1},q(Y_1,\narrowdots,Y_{n-1}),X_{k+1},\narrowdots,X_m).\]

For instance,
\texttt{p(X,Y),a(X),b(Y)} can be written also as
\texttt{p(a,Y),b(Y)} and then as \texttt{p(a,b)},
which is much more concise and looks like a standard term.

Also, we allow
\[
	p(X_1,\narrowdots,X_{k-1},L,X_{k+1},\narrowdots,X_m ),m\mem{\texttt{+}L,P}
\]
$(1 \leq k \leq m)$ to be written as
\[
	p(X_1,\narrowdots,X_{k-1},m\mem{P},X_{k+1},\narrowdots,X_m),
\]
where $\texttt{+}L$ is a operator notation for $\texttt{+}(L)$.

\subsection{Reduction Relation}\label{sec:reduction_relation}

\begin{figure}[tbp]
	\centering\small
	\begin{tabular}{llll}
		\hline                                                                                                                                                               \\[-2ex]
		(R1) & $\dfrac{P\reduces P'}{P,Q\reduces P',Q}\quad$
		     & (R2)                                                                                                   & $\dfrac{P\reduces P'}{m\{P\}\reduces m\{P'\}}\quad$~ \\[9pt]
		(R3) & \multicolumn{3}{l}{$\dfrac{Q\equiv P\hspace{10pt}P\reduces P'\hspace{10pt}P'\equiv Q'}{Q\reduces Q'}$}                                                        \\[9pt]
		(R4) & \multicolumn{3}{l}{$m\{X=Y,P\}\reduces X=Y,m\{P\}$}                                                                                                           \\
		     & \multicolumn{3}{r}{if $X$ and $Y$ are free links of $P$}                                                                                                      \\[1pt]
		(R5) & \multicolumn{3}{l}{$X=Y,m\{P\}\reduces m\{X=Y,P\}$}                                                                                                           \\[1pt]
		     & \multicolumn{3}{r}{if $X$ and $Y$ are free links of $P$}                                                                                                      \\[1pt]
		(R6) & \multicolumn{3}{l}{$T\theta,(T\texttt{:-}U)\reduces U\theta,(T\texttt{:-}U)$}                                                                                 \\[3pt]
		\hline
	\end{tabular}
	\caption{Reduction relation on LMNtal processes.}
	\vspace*{-10pt}
	\label{fig:Reduction relation on LMNtal processes}
\end{figure}

The reduction relation $\reduces$
is the minimum binary relation satisfying the rules
in \cref{fig:Reduction relation on LMNtal processes}.

(R1)--(R3) are structural rules.
%
(R4) and (R5) handle interaction between connectors and membranes,
and we leave the details to \cite{ueda_lmntal_2009}.
The key rule, (R6), handles rewriting (of the process $T\theta$)
by a rule.  The substitution $\theta$,
determined upon graph matching,
is to map process contexts (wildcards for non-rule processes),
rule contexts (wildcards for rewrite rules), and aggregates
into specific processes, rules, and collections of atoms,
respectively.  The precise formulation of
the substitution can be found in \cite{ueda_lmntal_2009}.

\qcenewU{%
	The reduction relation of QLMNtal is extended to handle both
	quantified processes and \textit{negative application condition}, an
	important feature for general graph rewriting that can be regarded as
	a special kind of quantifier \cite{mishina_introducing_2024}.
}

	\section{ZX-rules in plain LMNtal}\label{sec:zx-rules-in-plain-lmntal}
\begin{figure}[tb]
		\centering
		\includegraphics[width=\columnwidth]{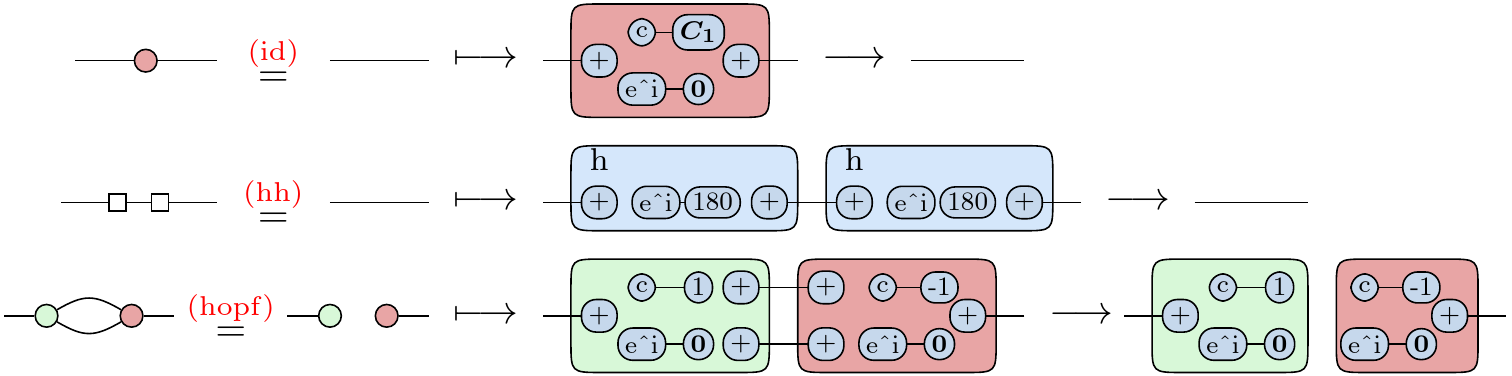}
		\vspace*{1em}
		\includegraphics[width=\columnwidth]{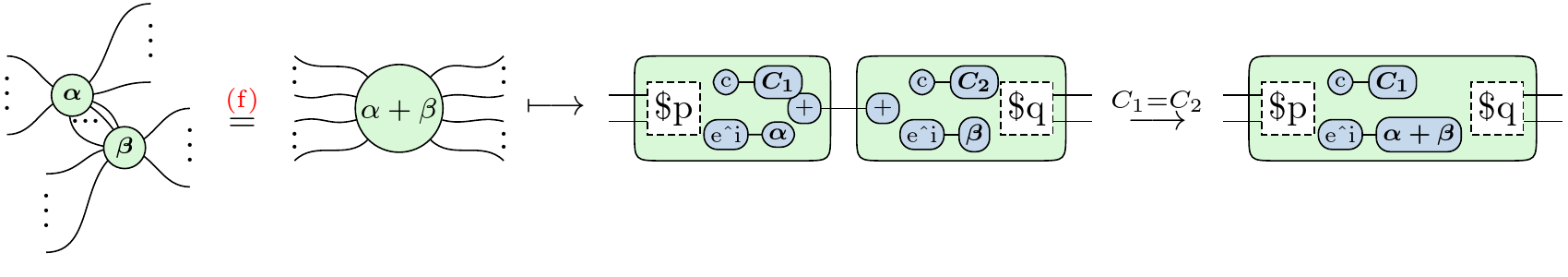}
		\vspace*{1em}
		\includegraphics[width=\columnwidth]{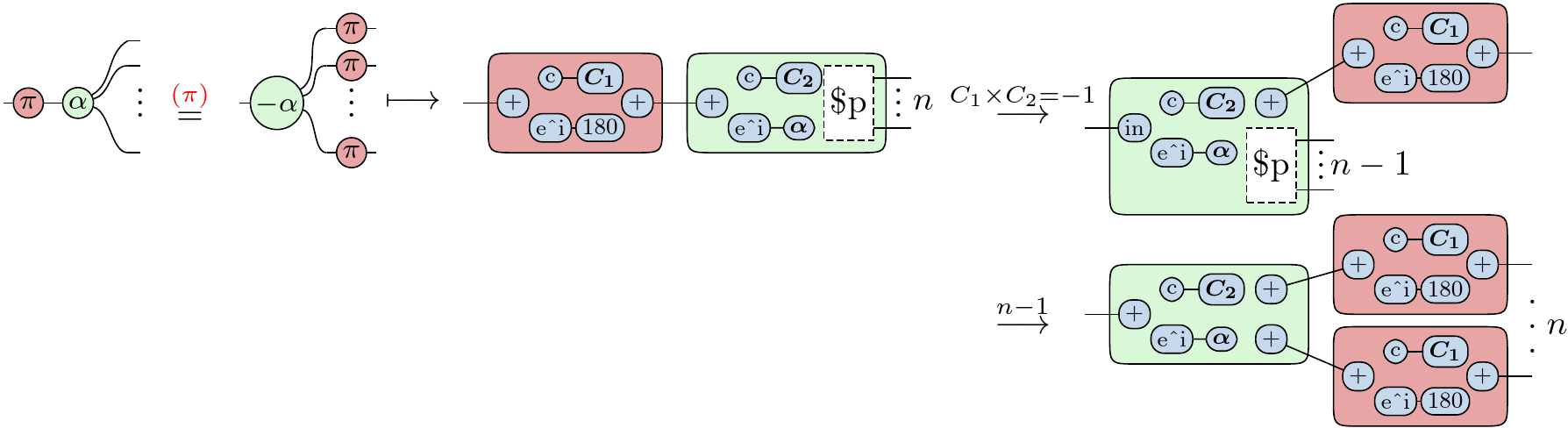}
		\vspace*{1em}
		\includegraphics[width=\columnwidth]{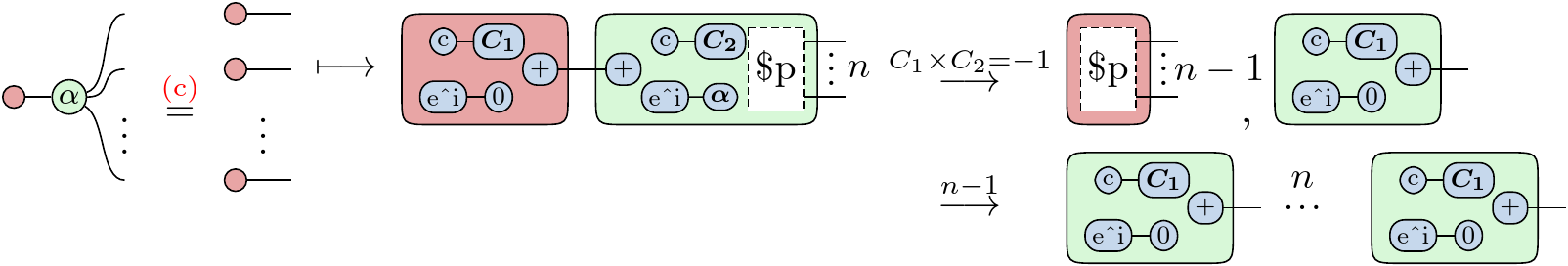}
		\vspace*{1em}
		\includegraphics[width=\columnwidth]{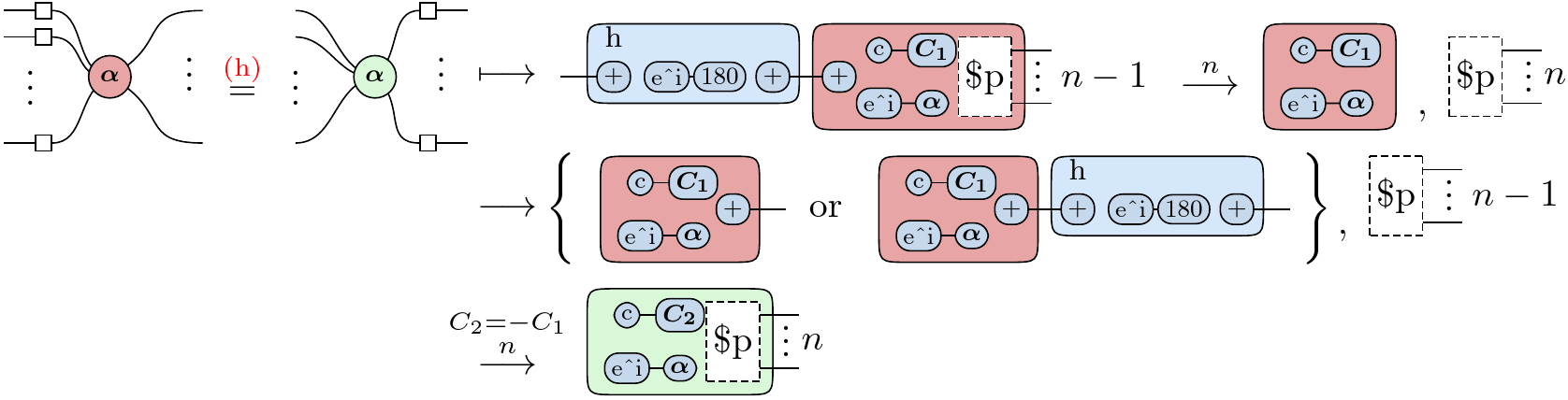}
		\vspace*{1em}
		\includegraphics[width=\columnwidth]{figs/b.png}
		\vspace*{1em}
		\caption{All the ZX-rules in plain LMNtal}
		\vspace*{-10pt}
		\label{fig:ZX-rules in plain LMNtal}
	\end{figure}

	In \cref{subsec:zx rule implementation}, we have shown how to implement {\color{red}(b)} rule in LMNtal. In the same way, all the other ZX-rules in \cref{fig:rules} can be implemented in plain LMNtal as shown in \cref{fig:ZX-rules in plain LMNtal}.

	\paragraph{Identity Removal, Hadamard Cancellation, Hopf:}
	Among the main rules of the ZX-calculus, {\color{red}(id)},
	{\color{red}(hh)}, and {\color{red}(h)} are the rules where the number
	of all wires appearing in the spiders is fixed.  Each of these rules can be represented by a single LMNtal rule.

	\paragraph{\color{black}Spider Fusion:}
	Operation {\color{red}(f)} combines spiders of the same color into one when they are connected by at least one wire.
	This rule can be expressed as a single LMNtal rule.
  If the spiders are connected by multiple wires, some remnant links
  remain inside the fused spider, and a minor additional rule is
  used to tidy them up.

	\paragraph{\color{black}$\pi$-commutation:}
	Operation {\color{red}($\pi$)} lets a spider with phase $\pi$ cross a spider
	of an opposite color and be possibly duplicated.
	This is handled by three LMNtal rules.
	The spider of phase $\pi$ is first moved to the output side of the
	spider of the other type, and then the spiders of phase $\pi$ are
	connected to all remaining free links of the other spider.
  In order to
  retain information about the wire to which the original
  $\pi$ spider was connected,
	the \texttt{+} atom is temporarily changed to an
	\texttt{in} atom.  When there are $n$ output wires,
	this rule takes $n+2$ steps overall.

	\paragraph{\color{black}Copy:}
	Operation {\color{red}(c)} duplicates a spider with phase 0 when it is
not connected to anything but a spider of an opposite color.
  {\color{red}(c)} can also be expressed in three LMNtal rules like
  {\color{red}($\pi$)}, and takes $n+2$ steps overall.

\paragraph{\color{black}Color Change:}
	Operation
  {\color{red}(h)} changes the color of a spider if one of the wires of a 	spider is directly connected to a Hadamard gate.
 Besides, all
Hadamard gates are removed from the wires, and Hadamard gates are
added to the wires without Hadamard gates.
	This approach combines the standard {\color{red}(h)} with {\color{red}(hh)} in \cref{fig:rules}.
	A direct implementation of the original {\color{red}(h)} rule is complex,
	because the number of Hadamard gates to match is unknown in advance
	and backtracking is required when it fails.
	In this version, matching succeeds and rewriting can be done
	if there is at least one Hadamard gate.
	{\color{red}(h)} consists of six LMNtal rules.
  Since this rule includes a conditional branch, it requires more
  pattern matching than {\color{red}($\pi$)} and {\color{red}(c)},
and takes $2n+2$ steps overall.

\paragraph{Generalized bialgebra:}
	Operation {\color{red}(b)} duplicates spiders as many times as the number of
	wires that are not connected between spiders of different colors, and
	connects the duplicated spiders to each other.
	This implementation consists of ten LMNtal rules.
  In order to duplicate all the spiders, we need to count the number of
  all the wires and then duplicate and connect them, which takes
  $mn+m+n$ steps.
  Including the preprocessing steps for duplication,
	it takes $mn+2m+n+3$ steps overall.

	\clearpage
	\section{ZH-Calculus}\label{sec:zh}
	The ZH-calculus \cite{backens2019zh}, an extension of the ZX-calculus,
	introduces a new generator called the H-box as in \cref{fig:hbox},
	which extends the input/output of the Hadamard gate to arbitrary
	numbers, enabling efficient representation of multi-controlled
	operations such as the Toffoli gate. H-boxes can naturally encode
	non-linear logical operations, making them well-suited for analyzing
	hybrid quantum-classical algorithms that incorporate classical logic
	gates.
	Among the basic rules of the ZH-calculus, this paper focuses only on
	the three rules in \cref{fig:ZH-rules in QLMNtal} (\cref{sec:impl})
	from \cite{vandewetering_zx-calculus_2020}.

	\begin{wrapstuff}[type = figure, r, width = 0.3\textwidth]
	\centering
	\includegraphics[width=\columnwidth]{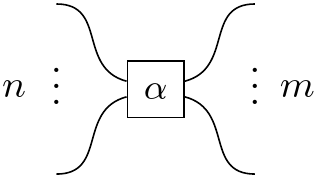}
	\caption{H-box\cite{vandewetering_zx-calculus_2020}.}
	\label{fig:hbox}
\end{wrapstuff}

	Even in the ZX-calculus,
	we encode Hadamard gates as 2-arity H-boxes with phase $\pi$
	because we want to extend ZX into ZH-calculus naturally.
	Membranes allow us to have arbitrary number of links, and
	the phase is always $\pi$ because
	an H-box (an extension of the
	Hadamard gate) with no label is treated as having phase $\pi$.

	The basic rules introduced in the ZH-Calculus can be
	represented using the H-box, which allows the number of links and the
	phase of the Hadamard gate to be arbitrarily changed.
	Although it is possible to implement all the basic rules in
	\cite{vandewetering_zx-calculus_2020}
	using plain LMNtal,
	\padl{\Cref{fig:ZH-rules in QLMNtal} shows
		three of the ZH-rules and their implementation in QLMNtal.}

	Although it is possible to implement all the basic rules in
	\cite{vandewetering_zx-calculus_2020}
	using \padl{plain} LMNtal,
	\padl{\Cref{fig:ZH-rules in QLMNtal} shows
		three of the ZH-rules and their implementation in QLMNtal.}

\begin{table}[tb]
	\centering
	\footnotesize
	\caption{QLMNtal implementation of ZH-rules}
	\begin{tabular}{c|l}
		\hline
		ZH-rule & \multicolumn{1}{c}{QLMNtal implementation}\\
		\hline
		\hline
		\begin{minipage}{0.25\textwidth}
			\centering
			\includegraphics[width=0.85\textwidth]{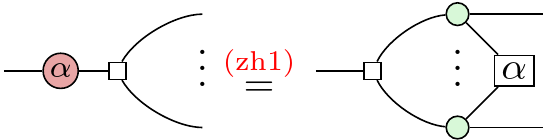}
		\end{minipage} &
		\begin{tabular}{@{}l@{}}
	\verb/{+L1,+L2,e^i(A),c(-1)}, h{+L2,<*>+L3,e^i(180)} :-/    \\
	\verb/  int(A) |/ \\
	\verb/  h{+L1,<*>+L2,e^i(180)}, h{<*>+L4,e^i(A)},/ \\
	\verb/  <*>{+L2,+L3,+L4,e^i(0),c(1)}./                      \\
		\end{tabular}
		\\
		\hline
		\begin{minipage}{0.25\textwidth}
			\centering
			\includegraphics[width=0.85\textwidth]{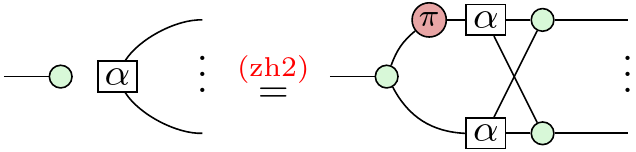}
		\end{minipage} &
		\begin{tabular}{@{}l@{}}
	\verb/{+L1,e^i(0),c(1)}, h{<*>+L2,e^i(A)} :- int(A) |/      \\
	\verb/  {+L1,+L3,+L4,e^i(0),c(1)}, {+L3,+L5,e^i(180),c(-1)},/ \\
	\verb/  <*>{+L2,+L6,+L7,e^i(0),c(1)},/                       \\
	\verb/  h{+L5,<*>+L6,e^i(A)}, h{+L4,<*>+L7,e^i(A)}./          \\
		\end{tabular}
		\\
		\hline
		\begin{minipage}{0.25\textwidth}
			\centering
			\includegraphics[width=0.85\textwidth]{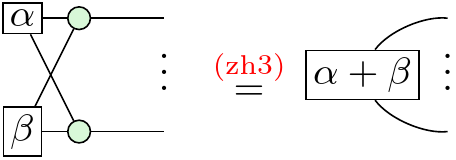}
		\end{minipage} &
		\begin{tabular}{@{}l@{}}
	\verb/<*>{+L1,+L2,+L3,e^i(0),c(1)},/                 \\
	\verb/h{<*>+L1,e^i(A)},h{<*>+L2,e^i(B)} :-/          \\
	\verb/  int(A), int(B), ApB=A+B | h{<*>+L3,e^i(ApB)}./ \\
		\end{tabular}
		\\
		\hline
	\end{tabular}
	\label{fig:ZH-rules in QLMNtal}
\end{table}

\clearpage
\padlY{
\section{Proof of Lemma A}
	For Lemma A
	\cite[Lemma 9.127]{picturing_quantum_2017}
	used in the proof of Lemma L in \cref{subsec:proof-lcomp}, we show the
	state space when proved with a similar manner in
	\cref{fig:lemma-a-state}.
	To search
	for our expected final state, we can use LTL model checking or
	visualization tools to check the resulting graph itself or
	consumption status of rule tokens.

\begin{figure}[t]
	\centering
	\includegraphics[width=0.8\columnwidth]{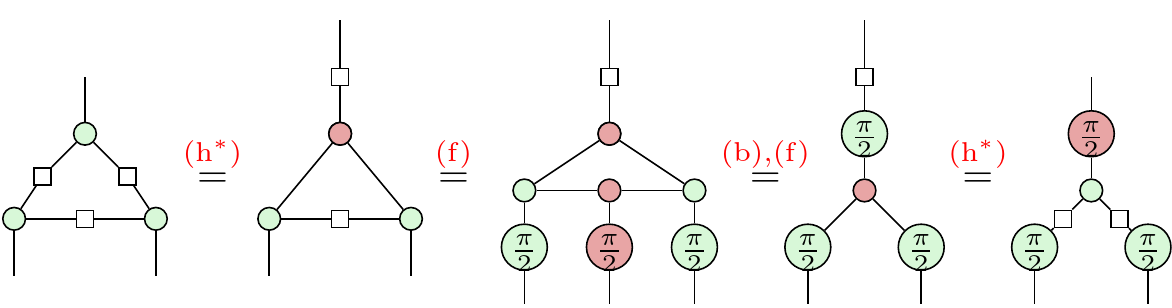}
	\caption{Proof Flow of Lemma A.}
	\vspace*{-10pt}
	\label{fig:lemma-a}
\end{figure}

\begin{figure}[t]
	\centering
	\includegraphics[width=\textwidth]{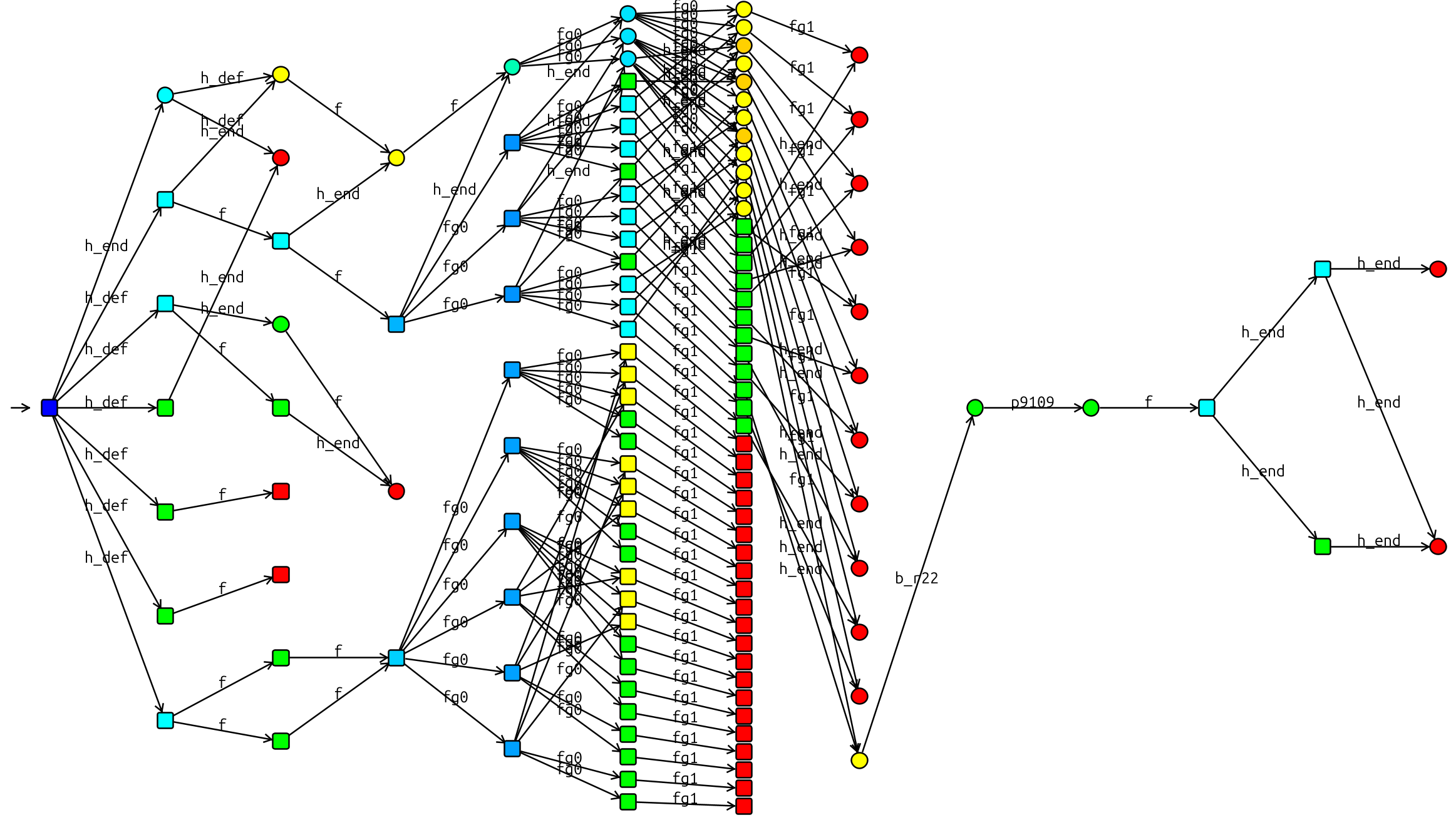}
	\caption{State Space for Lemma A (123 states).}
	\label{fig:lemma-a-state}
\end{figure}
}

\clearpage
\padl{
	\section{Analyzing State Space of 2-qubit QFT}\label{sec:qft}
	\Cref{fig:id_search} and \cref{fig:pi_search} show the rule search results in the state space
	when we analyze the 2-qubit QFT example in \cref{subsec:non-confluent}.
	Among all the final states shown in red,
	the longer the path to reach there, the smaller the number of spiders in the final state tends to be.
	That is, for who wants to simplify a given quantum circuit as much as possible,
	the longer paths are the better paths here.
	Comparing \cref{fig:id_search} and \cref{fig:pi_search},
	the number of applications of {\color{red}(id)} tends to be smaller
	in the paths to reach simpler final states.
	Moreover, {\color{red}($\pi$)} tends to be applied only in the middle part of the successful paths, not in the
	initial and final parts.
	This suggests that, to reach a simpler state, it is not effective to
	reduce or increase the number of spiders too much at the beginning.
	The above tendencies are only for the 2-qubit QFT example searched
	by our system, and it remains future work to see whether similar
	tendencies can be observed in other quantum circuits.

	\begin{figure}[t]
		\centering
		\includegraphics[width=.6\textwidth]{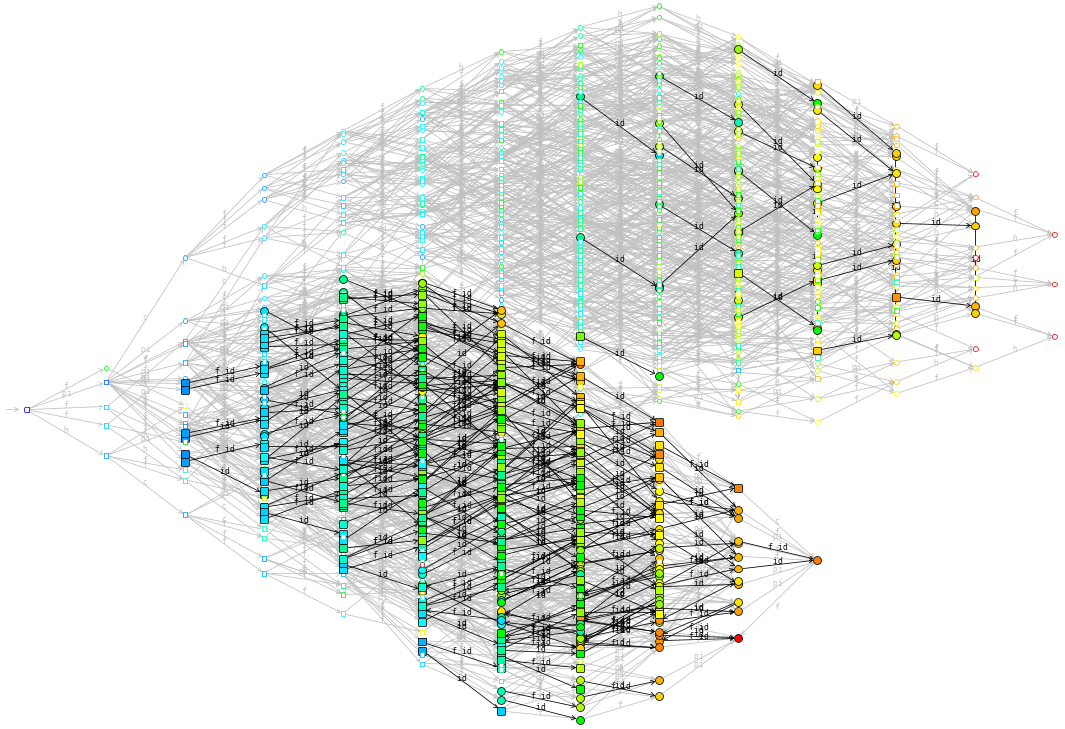}
		\caption{Searched {\color{red}(id)} in the 2-qubit QFT example}
		\label{fig:id_search}
	\end{figure}

	\begin{figure}[t]
		\centering
		\includegraphics[width=.6\textwidth]{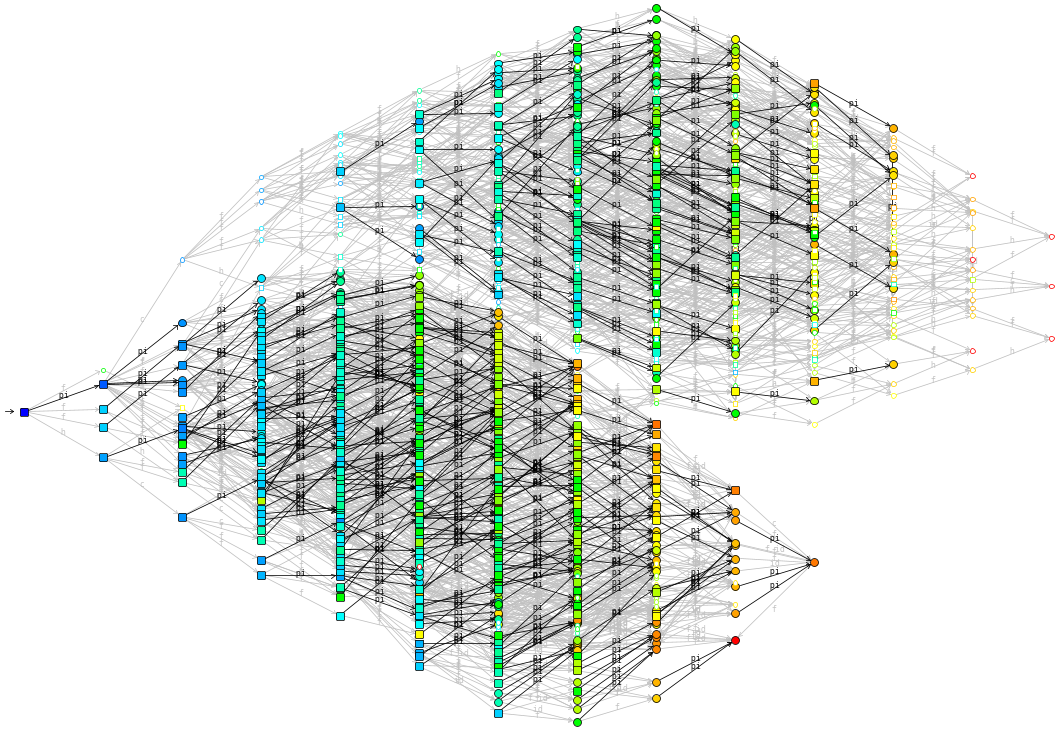}
		\caption{Searched {\color{red}($\pi$)} in the 2-qubit QFT example}
		\label{fig:pi_search}
	\end{figure}
}

\end{document}